\documentclass[11pt]{article} 
\usepackage{jheppub_no_preprint}

\usepackage{amsmath,amsfonts,amsbsy,amssymb,array,accents,dsfont}
\usepackage{enumerate}
\usepackage[shortlabels]{enumitem}
\usepackage{graphicx} 
\usepackage{subcaption} 
\usepackage{upgreek}
\usepackage{bm}
\usepackage{slashed}
\usepackage{tikz}
\usetikzlibrary{decorations.pathreplacing}
\let\includefigures=\iftrue
\let\useblackboard==\iftrue
\newfam\black

\usepackage[dvipsnames]{xcolor}

\definecolor{myblue}{RGB}{85,130,255}%{55, 100, 210}
\definecolor{myred}{RGB}{200, 45, 40}

\PassOptionsToPackage{ 
     colorlinks=true,
     linkcolor=myBlue,
     urlcolor=blue,
     filecolor=black,
     citecolor=myRed,
}{hyperref}

\usepackage{xparse}
\usepackage{xspace}

\NewDocumentCommand\eqn{om}{%
  \IfNoValueTF{#1}
     {\[ #2 \]}
     {\begin{equation}\label{#1} #2  \end{equation} \expandafter\newcommand\csname #1\endcsname{\eqref{#1}\xspace}\ignorespaces}
}
\NewDocumentCommand\eqna{om}{%
  \IfNoValueTF{#1}
    {\begin{align*} #2 \end{align*}}
    {\begin{equation}\label{#1}\begin{split} #2  \end{split}\end{equation} \expandafter\def\csname #1\endcsname{\eqref{#1}\xspace}\ignorespaces}
}

\def\sl{\text{sl}}

\def\({\left(}
\def\){\right)}
\def\[{\left[}
\def\]{\right]}

\def\sfI{{\mathsf I}}

\def\sfm{{\mathsf m}}

\DeclareMathSymbol{\medhatsym}{\mathord}{largesymbols}{"62} % basic symbol

\DeclareMathSymbol{\medtildesym}{\mathord}{largesymbols}{"65}% basic symbol

\makeatletter
\newcommand*\rel@kern[1]{\kern#1\dimexpr\macc@kerna}
\newcommand*\widebar[1]{%
  \begingroup
  \def\mathaccent##1##2{%
    \rel@kern{0.8}%
    \overline{\rel@kern{-0.8}\macc@nucleus\rel@kern{0.2}}%
    \rel@kern{-0.2}%
  }%
  \macc@depth\@ne
  \let\math@bgroup\@empty \let\math@egroup\macc@set@skewchar
  \mathsurround\z@ \frozen@everymath{\mathgroup\macc@group\relax}%
  \macc@set@skewchar\relax
  \let\mathaccentV\macc@nested@a
  \macc@nested@a\relax111{#1}%
  \endgroup
}
\makeatother

\def\One{{\hbox{1\kern-1mm l}}}

\def\Im{{\sfI\sfm\,}}

\def\barray{\begin{array}}
\def\earray{\end{array}}
\def\be{\begin{equation}}
\def\ee{\end{equation}}
\def\bea{\begin{eqnarray}}
\def\eea{\end{eqnarray}}
\def\bal{\begin{align}}
\def\eal{\end{align}}

\definecolor{cardinal}{rgb}{0.6,0,0}
\definecolor{darkgreen}{rgb}{0,0.4,0}
\definecolor{green}{rgb}{0,0.4,0}
\definecolor{golden}{rgb}{0.92, 0.7, 0}
\definecolor{midnight}{rgb}{0, 0, 0.5}
\definecolor{darkblue}{rgb}{0, 0, 0.7}

\numberwithin{equation}{section}

\mathchardef\mhyphen="2D

\def\one{{\hbox{\kern+.5mm 1\kern-.8mm l}}}
\def\zero{{\hbox{0\kern-1.5mm 0}}}

\def\id{\textrm{id}}

\def\id{{1 \kern-.28em {\rm l}}}

\def\journal#1&#2(#3){\unskip, \sl #1\ \bf #2 \rm(19#3) }
\def\andjournal#1&#2(#3){\sl #1~\bf #2 \rm (19#3) }

\def\One{{1\hskip -3pt {\rm l}}}

\catcode`\@=11
\def\slash#1{\mathord{\mathpalette\c@ncel{#1}}}
\def\underrel#1\over#2{\mathrel{\mathop{\kern\z@#1}\limits_{#2}}}

\catcode`\@=12

\def\exp{{\rm exp}}
\title{%\LARGE 
{
Deformed BTZ Radiance and Single Trace $T\bar{T}$ Holography
}}

\author{
Daniel Vainshtein
}

\affiliation{
\vskip 0.01cm
Racah Institute of Physics, The Hebrew University, Jerusalem, 91904, Israel
}

\emailAdd{daniel.vainshtein@mail.huji.ac.il 
}

\abstract{
We generalize the "holar wind" mechanism proposed in \cite{Martinec:2023plo} to the case of rotating $\lambda$-deformed BTZ black holes. These backgrounds,  which interpolate between $AdS_3$ in the infrared and a linear dilaton spacetime in the ultraviolet, are realized as an exact $\frac{SL(2,\mathbb{R})_k\times U(1)}{U(1)}$ gauged-WZW worldsheet theory. The long string sector of the theory provides a holographic dual to single-trace $T\overline{T}$ deformed symmetric product $\mathcal{M}^p/S_p$  $CFT_2$ with a seed $\mathcal{M}$ carrying a central charge $c=6k$. By analyzing the geodesics and tunneling rates of probe particles and winding long strings, we show that the emission probabilities for both positive and negative deformation couplings are universally governed by the change in the Bekenstein-Hawking entropy, $\Delta S_{BH}$. A central result of our analysis is that the consistency of long string emission within the grand canonical ensemble dictates a unique value for the background B-field at the origin. We demonstrate that this thermodynamically fixed value matches precisely the prediction required for the string excitation spectrum to agree with the $\mathbb{Z}_w$ twisted sector of a single-trace $T\overline{T}$ deformed symmetric product orbifold. Finally, we sketch the extension of these results to a broader class of black holes whose long string sector is dual to single-trace $T\overline{T} + J\overline{T} + T\overline{J}$ deformed theories.
}

\begin{document}
\hypersetup{pageanchor=false}
\begin{titlepage}
\maketitle
\thispagestyle{empty}
\end{titlepage}
\hypersetup{pageanchor=true}
\pagenumbering{arabic}

%\toc
\thispagestyle{empty}

%\vskip 1cm
%\hrule

%\begin{enumerate}[start=1,
%    labelindent=\parindent,
%    leftmargin =2.5\parindent,
%    label=(WII-\arabic*)]
%\item
%\label{WII-1}
%stuff
%\item
%\label{WII-2}
%more stuff
%\end{enumerate}

%Later on I want to refer to \ref{WII-1}

\section{Introduction} 
\label{sec:intro}
The $AdS_3/CFT_2$ correspondence provides one of the most tractable and well-understood examples of holographic duality. It is realized through string theory on $AdS_3$ backgrounds supported by an NS-NS $B$-field flux, constructed by near-$k$ NS5 branes wrapping $S^1\times \mathbb{T}^4$ with $p$ fundamental strings winding $S^1$. The corresponding worldsheet theory becomes exactly solvable as a WZW model on $SL(2,\mathbb{R})_k$. This solvability allows for a detailed investigation of the string spectrum and its holographic interpretation \cite{Giveon:1998ns,Kutasov:1999xu,Maldacena:2000hw}.
A particularly interesting sector consists of long string states, which are delta-function normalizable states carrying winding number $w$ that extend throughout the asymptotic $AdS_3$ region. The spectrum of these long strings exhibits the structure of an untwisted sector for $w=1$ and $
\mathbb{Z}_w$ twisted sector for $w>1$ of a $p$-fold symmetric product, $\mathcal{M}^p/S_p$, where the seed $\mathcal{M}$ is a $CFT_2$ with central charge $c = 6k$ describing the dynamics of a single long string 
\cite{Argurio:2000tb,Seiberg:1999xz}. This structure persists remarkably under certain exactly marginal deformations of the worldsheet theory.
One such deformation, studied extensively in recent years, is characterized in the bulk theory as interpolating between $AdS_3$ in the infrared and an asymptotically flat spacetime with linear dilaton in the ultraviolet \cite{Asrat:2017tzd,Giveon:2017myj}. From the worldsheet perspective, this amounts to a truly marginal deformation of the $SL(2,\mathbb{R})_k$ WZW model. In the dual two-dimensional field theory, it corresponds to an irrelevant deformation which shares many properties of a $T\bar{T}$ deformation \cite{Giveon:2017nie}. The resulting framework, termed "single-trace $T\bar{T}$ holography" is characterized by the structure of a p-fold symmetric product
of a $T\bar{T}$ deformed CFT seed $\mathcal{M}$.
Such theories has proven to be a rich testing ground for the emergence of non-locality in quantum field theory \cite{Smirnov:2016lqw,Cavaglia:2016oda,Giribet:2017imm}, complexity and entanglement entropy \cite{Chen:2018eqk,Chakraborty:2018kpr,Chakraborty:2020udr,Chakraborty:2025rfm} (For a review and the current status of $T\bar{T}$ deformation, see \cite{He:2025ppz}).
Another family of solvable irrelevant deformations of the boundary $CFT_2$ are include the operation of  $T\bar{J},J\bar{T}$ operators which constructed by additional conserved left-right $J,\bar{J}$ currents. Their study was initiated in \cite{Guica:2017lia,Chakraborty:2018vja,LeFloch:2019rut} and later shown to describe deformations on $AdS_3\times S^1$ obtained from a truly marginal deformations on the extended $SL(2,\mathbb{R})_k\times U(1)$ worldsheet theory \cite{Giveon:2019fgr,Chakraborty:2019mdf}.
This similar framework, termed "single-trace $T\bar{T}+J\bar{T}+T\bar{J}$ holography" corresponds to the structure of a p-fold symmetric product, with a CFT seed $\mathcal{M}$ deformed by a 
 linear combination of $T\bar{T},T\bar{J},J\bar{T}$ operators\footnote{As in the $T\bar{T}$ case, single trace $T\bar{T}+J\bar{T}+T\bar{J}$ constitute a rich tool to test complexity and entanglement entropy as well 
\cite{Chakraborty:2020udr,Katoch:2022hdf}.}.

A central use of the holographic duality lies in the exploration of black hole physics, which constitutes a primary example for investigating the thermodynamics of finite temperature states.
In the context of the single trace $T\bar{T}$ in deformed $AdS_3/CFT_{2}$ holography, the bulk theory solutions are $\lambda$-deformed BTZ black holes, with a dimensionless coupling $\lambda$ which can be either positive $\lambda>0$ or negative $\lambda<0$. The study of the massless case of these solutions was initiated in \cite{Giveon:2017nie} and later generalized for general black hole backgrounds carrying mass and angular momentum \cite{Apolo:2019zai,Chakraborty:2020swe,Chakraborty:2023mzc,Chang:2023kkq,Chakraborty:2023zdd}.
These solutions correspond to a high energy states of the Ramond-Ramond sector, such that  energy, momentum and entropy split equally among the $p$ copies of the symmetric orbifold $\mathcal{M}^p/S_p$. 
In particular, the ground state of the NS-NS sector in the deformed $T\bar{T}$ theory was found to be compatible of a $\lambda$-deformed global $AdS_3$ on the bulk.
An important question in this context concerns the mechanism of Hawking radiation from these black holes. In the standard undeformed BTZ black holes, massive particle excitations that are emitted from a black hole cannot escape to infinity; they reach a maximum radius and fall back. Massless quanta can reach the conformal boundary but are reflected and refocused back onto the black hole. Thus, there is no S-matrix for particles in $AdS_3$; observables are instead defined as insertions along the timelike conformal boundary and can model BTZ black hole decay. An alternative approach, which will be demonstrated in this note, considers backgrounds in which the AdS throat open up to an asymptotically flat regime at radius $R$. 

When considering winding string excitations in the theory, carrying winding number $w \ge 1$ above a certain energy threshold lie in plane-wave normalizable states that can propagate outward a timelike infinity at the spatial boundary \cite{Hemming:2002kd,Troost:2002wk}. These states form a long string continuum and provide a genuine decay channel for BTZ black holes through the emission of wound fundamental strings. This phenomenon, described in \cite{Martinec:2023plo} as the "holar wind" (by analogy with the solar wind) represents a black hole S-matrix within BTZ black holes.

In this note, we extend this analysis to the case of $
\lambda$-deformed rotating BTZ black holes. 
In section \ref{sec:positive}, we begin by reviewing the properties of the black-string solution for positive coupling $
\lambda>0$, examining its compatibility with single trace $T\bar{T}$ holography with respect to maximally twisted sector of the symmetric orbifold \cite{Chakraborty:2023zdd}. 
For the excited string states, the spectrum can be extracted by using only the asymptotic data, thus, relying on recent work \cite{Chakraborty:2024ugc,Giveon:2024sgz}, we
predict a non-trivial value for the $B$-field in the origin $B_{\tau\theta}^{(0)}$,which is required for rotating deformed solutions to maintain "harmony with single trace $T\bar{T}$ holography".

In \ref{sec:pos_particle}-\ref{2.4}, we present the dynamics of probe particles and long strings in the presence of the deformed black hole. We first derive their physical geodesics, and then performing a semiclassical calculation for the emission amplitude of a quantum from the black hole using the tunneling and WKB paradigms. 
For particle dynamics the black hole background treated as closed thermodynamical atmosphere within the microcanonical ensemble. Conversely, for long string emission, the picture extended to treat fluctuations in the total winding number, necessitating the use of the grand canonical ensemble.
Taking this thermodynamical pictures into account, we find that for linear dilaton black holes, the emission probability still obeys the universal result \eqref{Gamma2} for both particles and strings. That is, the probability is decays exponentially with the change in the Bekenstein-Hawking entropy during the emission process. This makes the claim stated for such backgrounds in \cite{Martinec:2023plo} explicit.
As a direct consequence of this result, we find that black hole thermodynamics dictate a unique, favored value for $B_{\tau\theta}^{(0)}$ that precisely matches the value predicted by single trace $T\bar{T}$ holography.  

In section \ref{sec:negative},
we repeat the calculation for negative deformation coupling.
This case is particularly interesting as it involves "negative strings" \cite{Chakraborty:2020swe} and exhibits additional features such as a maximum black hole and strings energy, emergence of pathologies such as close time curves (CTCs) and a UV cutoff singularity.
Thus, modifying entirely the dynamics of emitted particle and string from that of a positive deformed smooth geometry.

In section \ref{sec:exten_TJJT}, we sketch the generalization of long string and particle emission to a wider class of black hole solutions, which are similarly characterized by asymptotic flatness with linear dilaton. These are $(\lambda,\epsilon_{\pm})$-deformed BTZ$\times S^1$ black holes, whose energy spectrum matches that of a maximally twisted $\mathbb{Z}_p$ sector of a single trace $T\bar{T}+J\bar{T}+T\bar{J}$ deformed symmetric orbifold $\mathcal{M}^p/S_p$. In this background, excited probe long strings carrying winding $w$ reside in the partially twisted $\mathbb{Z}_w$ sector.  
In more general setup \cite{Apolo:2021wcn},
these black holes may carry extra left and right charges,
which further modify the first law of thermodynamics and impose constraints on the additional chemical potentials.

We conclude in section \ref{sec:discussions} by discussing the broader consequences of our results. We emphasize the thermodynamic uniqueness of the choice of the $B$-field at the origin in the long string emission process, and compare our findings with different claims in the literature.
 Following  \cite{Martinec:2023plo}, we also summarize the various proposals for preserving unitarity during black hole evolution, and suggest future studies to analyze the behavior of the Hawking radiated particles and strings as seen by a distant observer.

\section{Positive Deformation}
\label{sec:positive}
Consider superstring theory on the deformed BTZ black holes background formed by near-$k$ NS5 branes wrapping $S^1\times \mathbb{T}^4 $ with $p$ fundamental strings (F1) and winding  $S^1$, whose radius we denote by $R$.
The strings carry quantized momentum number $n\in \mathbb{Z}$.
The deformed BTZ background, as presented in \cite{Chakraborty:2023zdd}, is given by:
\begin{equation}
\label{defbtz}
ds^2=-{N^2\over 1+{\rho^2\over R^2}}d\tau^2+{d\rho^2\over N^2}+{\rho^2\over 1+{\rho^2\over R^2}}(d\theta-N_\theta d\tau)^2,\qquad \theta\simeq\theta+2\pi
\end{equation}
\begin{equation}
\label{btautheta}
B_{\tau\theta}={\rho^2\over r_5}\sqrt{\left(1+{\rho_-^2\over R^2}\right)\left(1+{\rho_+^2\over R^2}\right)}{1\over 1+{\rho^2\over R^2}}+B_{\tau\theta}^{(0)}~,
\qquad B_{\tau\theta}^{(0)}\equiv B_{\tau\theta}(\rho=0)
\end{equation}
\begin{equation}
\label{dilaton}
e^{2\Phi}={kv\over p}\sqrt{\left(1+{\rho_-^2\over R^2}\right)\left(1+{\rho_+^2\over R^2}\right)}{1\over 1+{\rho^2\over R^2}}~,
\qquad v\equiv {\rm Volume}(\mathbb{T}^4)/\left(2\pi\sqrt{\alpha'}\right)^4
\end{equation}
with

\begin{equation}
\label{withn}
N^2={\left(\rho^2-\rho_+^2\right)\left(\rho^2-\rho_-^2\right)\over r_5^2\rho^2}, \qquad N_\theta={\rho_+\rho_-\over r_5\rho^2}, \qquad r_5\equiv\sqrt{\alpha' k}.
\end{equation}

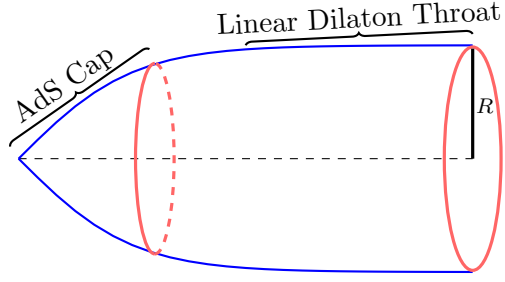
\begin{figure}[h!] 
    \centering
    \begin{tikzpicture}
[scale = 1.5,domain=0:4]
  %\draw[ thin,color=gray] (-0.1,-1.1) grid (3.9,3.9);

  %\draw[->] (-0.2,0) -- (4.2,0) node[right] {$x$};
  %\draw[->] (0,-1.2) -- (0,4.2) node[above] {$f(x)$};
  
% deformed positive case

  \draw[decorate,decoration={brace}, thick,rotate = 35] (0,0.13)--(1.5,0.13) node[rotate = 35] at (0.7,0.3)  {AdS Cap} ;
  
 \draw[decorate,decoration={brace}, thick] (2,1.0)--(4,1.1) node[rotate = 3] at (3,1.25)  {Linear Dilaton Throat} ;
  
  \draw[thin,dashed,color=black]    plot (\x,0)              ;
  % \x r means to convert '\x' from degrees to _r_adians:
  \draw[thick,color=blue]   plot (\x,{sqrt((sinh(\x))^2/(1+(sinh(\x))^2))})   ;
  \draw[thick,color=blue]   plot (\x,{-sqrt((sinh(\x))^2/(1+(sinh(\x))^2))}) ;
  
  %\draw[color=red!60 , very thick](1.5,0) ellipse (0.17 and 0.69230769230); %Full ellipse
   % y = 0.768221279597
   \draw[color=red!60 , very thick] (1.2,0.84) arc [
      start angle = 90,
      end angle = 270,
      x radius= 0.17 ,
      y radius = 0.84 
  ];
  \draw[color=red!60 , very thick, dashed] (1.2,-0.84) arc [
      start angle = 270,
      end angle = 450,
      x radius= 0.17 ,
      y radius = 0.84 
  ];
  
  \draw[color=red!60, very thick](4,0) ellipse (0.25 and 0.985);

  \draw[black, very thick] (4,0) -- (4,0.970142500145);
  \node[right] at (3.94,0.47055) {\scriptsize $R$};
  
  % undeformed case 
  %\draw[thick,color=orange,domain=0:1.2]   plot (\x,{(\x)});
   %\draw[thick,color=orange,domain=1.2:1.6,dashed]   plot (\x,{\x});
   
  %\draw[thick,color=orange,domain=0:1.2]   plot (\x,{-(\x)}); 
  %\draw[thick,color=orange,domain=1.2:1.6,dashed]   plot (\x,{-(\x)});
  
  %\draw[color=green!60 , very thick](1.2,0) ellipse (0.17 and 1.2);
\end{tikzpicture}  
    \caption{Positive $\lambda$-deformed global $AdS_3$ spatial geometry: for small $\rho$, the geometry is described by the $AdS_3$ cap which is smoothly widens with $\rho$ towards an asymptotically flat throat regime with asymptotic radius $R$. The metric is accompanied by a dilaton field that becomes linear in the asymptotic regime, supported by a constant $B$-field.}
    \label{fig1}
\end{figure}

This background describes a rotating black string in $2+1$ dimensions, where the B-field is defined up to a constant shift $B_{\tau\theta}^{(0)}$, representing the value of the B-field at the origin $B_{\tau\theta}(\rho =0)$. The geometry of the solution interpolates smoothly between an $AdS_3$ space in the IR and a flat throat in the UV. The geometry of the positively deformed global $AdS_3$ (amounts to $\rho_{-}=0,\rho_+=ir_5$) is depicted in Figure \ref{fig1}.
One can see this explicitly, by taking the limit $\frac{\rho}{R}\to \infty$ which constitutes the limit where the long string dominate. Then, transforming to canonical normalized coordinates: 
\begin{equation} \label{canon_norm}
 t = \frac{R}{r_5}\tau,\qquad x=R\theta,\qquad \phi = r_5\log\left(\frac{\rho}{R}\right) 
\end{equation}
yields that the solution \eqref{defbtz}-\eqref{dilaton} reduces exactly to a flat spacetime metric with linear dilaton and constant B-field at the asymptotics:
\begin{equation} \label{Fmetric}
ds^2 = -dt^2 + d\phi^2 + dx^2,\qquad x\sim x + 2\pi R
\end{equation}
\begin{equation} \label{FBdialaton}
\Phi = \Phi_0 - \frac{\phi}{r_5},\qquad B_{tx}   = 
\sqrt{\left(1+{\rho_-^2\over R^2}\right)\left(1+{\rho_+^2\over R^2}\right)}+B_{tx}^{(0)}.
\end{equation}

The black hole carries an ADM mass and angular momentum that can be expressed in terms of the inner and outer horizons $\rho_\pm$ and the asymptotic radius $R$:
\begin{equation} \label{M,J}
  M_{BH}=\frac{Rp}{\alpha'}\frac{1+\frac{\rho_{-}^{2}+\rho_{+}^{2}}{R^{2}}}{\sqrt{\left(1+\frac{\rho_{-}^{2}}{R^{2}}\right)\left(1+\frac{\rho_{+}^{2}}{R^{2}}\right)}},\qquad J_{BH}=\frac{p}{\alpha'}\frac{\rho_{-}\rho_{+}}{\sqrt{\left(1+\frac{\rho_{-}^{2}}{R^{2}}\right)\left(1+\frac{\rho_{+}^{2}}{R^{2}}\right)}}.
\end{equation}
The energy of the black hole is defined as the energy above extremality bound of the NS5-F1 black string, such that the extremal limit yields a zero ground-state energy\footnote{See e.g. eq.(2.10) in \cite{Chakraborty:2020swe} and eq.(2.8) in \cite{Chakraborty:2023zdd}}:
\begin{equation} \label{E_BH+}
E_{BH} = M_{BH} - E_{ext} =   \frac{Rp}{\alpha'} \left(-1 +\frac{1+\frac{\rho_{-}^{2}+\rho_{+}^{2}}{R^{2}}}{\sqrt{\left(1+\frac{\rho_{-}^{2}}{R^{2}}\right)\left(1+\frac{\rho_{+}^{2}}{R^{2}}\right)}} \right).
\end{equation}

Other thermodynamic properties, including the entropy, temperature and angular velocity, are given by:
\begin{equation} \label{S,T,OM}
S_{BH}=\frac{2\pi kp}{r_{5}}\frac{\rho_{+}}{\sqrt{1+\frac{\rho_{-}^{2}}{R^{2}}}},\qquad T_{BH}=\frac{\rho_{+}^{2}-\rho_{-}^{2}}{2\pi r_{5}R\rho_{+}\sqrt{1+\frac{\rho_{+}^{2}}{R^{2}}}},\qquad\Omega_{BH}=\frac{\rho_{-}}{R\rho_{+}}.
\end{equation}
When the system is found in the microcanonical ensemble with fixed $p,k$, and for fixed $R$, the following  first law of black hole thermodynamics is satisfied:
\begin{equation} \label{First Law}
dS_{BH}=\frac{1}{T_{BH}}\left(dE_{BH}-\Omega_{BH}dJ_{BH}\right).
\end{equation}
We define the \textit{positive} deformation parameter in terms of the asymptotic radius $R$ as:
\begin{equation} \label{lambda def}
 \lambda = \frac{\alpha'}{R^2}>0.
\end{equation}
 The limit $\lambda \to 0$ (or $R/\sqrt{\alpha'} \to \infty$) describes the undeformed limit, in which the solution \eqref{defbtz}- \eqref{dilaton} reduces to the standard BTZ solution\footnote{In \cite{Martinec:2023plo} the following coordinates and conventions were used: 
 \begin{equation*}
 r=\rho,\quad r_{\pm}=\rho_{\pm},\quad t = \frac{\tau}{r_{5}^2},\quad \phi = \frac{\theta}{r_5},\quad \ell = r_5  
 \end{equation*}}~\footnote{The integration constant in going from $H$ to $B$ is not ambiguous; as in~\cite{Chakraborty:2023mzc,Chakraborty:2023zdd}, it is fixed such that the $B$-field in~(\ref{btautheta}) for the BTZ limit vanishes at $\rho=0$. This is compatible with the choice argued for in~\cite{Ashok:2021ffx,Martinec:2023plo}.}:
\begin{equation}\label{btxmetric}
\begin{aligned}
   &  ds^2=-N^2d\tau^2+\frac{d\rho^2}{N^2}+\rho^2(d\theta-N_{\theta}d\tau)^2~,\\
 &  B_{\tau\theta}=\frac{\rho^2}{r_5} ~,\qquad
   e^{2\Phi}=\frac{kv}{p}~
\end{aligned}
\end{equation}
with $N^2,N_{\theta}$ given in \eqref{withn}.
In theories where the angular momentum $J_{BH}$ and the entropy $S_{BH}$ are fixed along the deformation line  parametrized by $\lambda$, i.e:
\begin{equation} \label{fixSJ}
  S_{BH}(\lambda)=S_{BH}(0)\equiv S_{BTZ} = \frac{2\pi kp}{r_5}\tilde{\rho}_+,\qquad J_{BH}(\lambda) = J_{BH}(0) \equiv J_{BTZ} = \frac{p}{\alpha'}\tilde{\rho}_-\tilde{\rho}_+
\end{equation}
where the fixed horizons $\tilde{\rho}_\pm$ are given in terms of $\rho_\pm$:
\begin{equation}
 \tilde\rho_\pm\equiv\frac{\rho_\pm}{\sqrt{1+\frac{\rho_\mp^2}{R^2}}},\qquad \tilde{\rho}_{\pm}(\lambda) =  \tilde{\rho}_{\pm}(0) = \rho_{\pm}(0),
\end{equation}
the energy of the deformed BTZ black hole coincides with that of the maximally twisted sector $\mathbb{Z}_p$ of a single trace $T\bar{T}$ deformed symmetric orbifold $\mathcal{M}^p/S_p$, where each deformed seed $\mathcal{M}$ carrying $1/p$ of the total energy:
\begin{equation} \label{pos_TT_energy}
 E_{BH}(\lambda)/p=\frac{1}{\lambda R}\left(-1+\sqrt{1+2\lambda R\frac{E_{BH}(0)}{p}+\left(\frac{\lambda RP_{BH}}{p}\right)^2}\right)
\end{equation}
where the undeformed energy and momentum are given by:

\begin{equation}
E_{BH}(0) = \frac{r_5}{R} M_{BTZ}=\frac{r_5p}{R} \frac{\tilde{\rho}_+^2+\tilde{\rho}_-^2}{2\alpha'},\qquad P_{BH} = \frac{J_{BTZ}}{R} = \frac{p}{R} \frac{\tilde{\rho}_-\tilde{\rho}_+}{\alpha'}.
\end{equation}
The total entropy, for large $R$ can be expressed in the Cardy form:

\begin{equation} \label{Entropy}
S_{BH} = 2\pi\sqrt{\frac{c}{6}}\left(\sqrt{E_{L}\left(1+\frac{\lambda}{p}E_{R}\right)}+\sqrt{E_{R}\left(1+\frac{\lambda}{p}E_{L}\right)}\right),\qquad c = 6kp 
\end{equation}
where the left and right energies are given by:
\begin{equation} \label{ELR}
E_{L,R}(\lambda)=\frac{1}{2}\left(E_{BH}\left(\lambda\right)R\pm J_{BH}(\lambda)\right).
\end{equation}
In particular, by identifying the entropy \eqref{Entropy} and the angular momentum with the undeformed BTZ quantities as was done in \eqref{fixSJ} one gets the following flow equation:
\begin{equation}
E_{L,R}(\lambda)+\frac{\lambda}{p}E_{L}(\lambda)E_{R}(\lambda) = E_{L,R}(0),\qquad J_{BH}(\lambda) = J_{BH}(0)=J_{BTZ}
\end{equation}
which, for the positive branch, yields the deformed energy spectrum \eqref{pos_TT_energy}.
By looking at the p's fraction of the total entropy \eqref{Entropy}, one can extract the undeformed seed $\mathcal{M}$ central charge, which is $c=6k$. 

In the following sections, utilizing the definitions above, we will calculate the geodesics and emission rate from a deformed BTZ background within the semiclassical tunneling and WKB approaches, considering the worldsheet sigma model action:
\begin{equation} \label{I_WS}
I_{WS} = -\frac{1}{4\pi \alpha'} \int d^2z\sqrt{\gamma} \left[\left(\gamma^{ab}G_{\mu\nu} + \varepsilon^{ab}B_{\mu\nu}\right)\partial_a x^\mu \partial_{b}x^\nu +\alpha'\Phi(x)R^{(2)}\right].
\end{equation}
We begin by analyzing the particle emission dynamics and then generalizing the analysis to a winding string dynamics, following the approach of \cite{Martinec:2023plo}.
In the upcoming analysis, we utilize Weyl invariance and and conveniently fix the worldsheet metric to be:
\begin{equation} \label{WS_metric}
d\gamma ^2 = -\frac{1}{r_5^2}d\xi^2+d\sigma^2,\qquad \sigma \sim \sigma + 2\pi
\end{equation}
The analysis for a general $\gamma^{ab}$ worldsheet metric is presented in Appendix \ref{appA}.

\subsection{Particle Geodesics}
\label{sec:pos_particle}
We consider the particle limit of the worldsheet action by integrating out the stringy degree of freedom. For the worldsheet metric in \eqref{WS_metric}, this yields the worldline action\footnote{In the particle case, we normalize the action with respect to the level $k$}:
\begin{equation} \label{worldline}
I_{ptcl} = \frac{1}{2r_5}\int d\xi \left(G_{\mu\nu}\dot{x}^\mu\dot{x}^{\nu}-r_5^2m^2\right)
\end{equation}
where $m^2$ represents the particles mass squared. The particle follows null trajectories for $m^2=0$, timelike and spacelike trajectories for $m^2>0,m^2<0$ respectively.
For the particles emission dynamics, potential effects coming from the B-field and the dilaton field are suppressed, leaving the metric $G_{\mu\nu}$ to completely dictate the geodesics.
The geodesics are determined by the conversation laws for a particle carrying energy $E$ and angular momentum $L$:
\begin{equation}  \label{Eptcl}
E=-p_{t}=G_{\mu\nu}k_{t}^{\mu}u^{\nu}=\frac{1}{r_{5}\left(1+\frac{\rho^{2}}{R^{2}}\right)}\left(\frac{\rho^{2}-\rho_{+}^{2}-\rho_{-}^{2}}{R^{2}}\dot{t}+\frac{\rho_{-}\rho_{+}}{R}\dot{\theta}\right)
\end{equation}
\begin{equation} \label{Lptcl}  L=p_{\theta}=G_{\mu\nu}k_{\theta}^{\mu}u^{\nu}=\frac{1}{r_{5}\left(1+\frac{\rho^{2}}{R^{2}}\right)}\left(\rho^{2}\dot{\theta}-\frac{\rho_{-}\rho_{+}}{R}\dot{t}\right).
\end{equation}
Here, $E,L$ are the physical quantities measured by an observer in the asymptotically flat regime with respect to the canonically normalized time $t$ \eqref{canon_norm}, while for simplicity we keep $\theta$ as the standard angular coordinate, and $\rho$ as the radial coordinate.
In addition to these conserved quantities, the Hamiltonian constraint on the mass squared of the particle must be satisfied, taking the form:
\begin{equation} \label{H_constr}
    \mathcal{H}_{ptcl} \propto G^{\mu\nu}p_\mu p_\nu +m^2=0
\end{equation}
where $p_{\mu}$ is the canonical covariant momentum conjugate to $x^{\mu}$, given explicitly by:
\begin{equation} \label{p_mu}
p_{\mu}=\frac{\partial\mathcal{L}_{ptcl}}{\partial{\dot{x}^{\mu}}}=\frac{1}{r_5}G_{\mu\nu}\dot{x}^\nu
\end{equation}
Substituting \eqref{Eptcl}-\eqref{Lptcl} and \eqref{p_mu} into the constraint \eqref{H_constr}, we obtain an explicit differential equation for the radial geodesics:
\begin{equation} \label{eq_rptcl}
    \dot{y}^2=4m_{\text{eff}}^2\left(-y^2+\frac{\alpha_d}{m_{\text{eff}}^2} y+\frac{\gamma_d}{m_{\text{eff}}^2}\right)
\end{equation}
where we define:
\begin{equation} \label{meff}
    y=\rho^2,\quad m_{\text{eff}}^{2}=m^{2}-\frac{E^{2}R^{2}-L^{2}}{R^{2}}
\end{equation}
along with the parameters,
\begin{equation}
    \alpha_{d}(m_{\text{eff}}^2)=E^{2}R^{2}-L^{2}+m_{\text{eff}}^{2}\left(\rho_{+}^{2}+\rho_{-}^{2}\right)+\frac{E^{2}R^{2}\left(\rho_{+}^{2}+\rho_{-}^{2}\right)-2\rho_{-}\rho_{+}ERL}{R^{2}}
\end{equation}
\begin{equation}
    \gamma_{d}(m_{\text{eff}}^2)=\left(\rho_{+}^{2}+\rho_{-}^{2}\right)L^{2}-2ERL\rho_{-}\rho_{+}-m_{\text{eff}}^{2}\rho_{+}^{2}\rho_{-}^{2}-\frac{E^{2}R^{2}-L^{2}}{R^{2}}\rho_{+}^{2}\rho_{-}^{2}.
\end{equation}
Equation \eqref{eq_rptcl} can be solved analytically, yielding the following physical solution across the different regimes of the effective mass squared $m_{\text{eff}}^2$:
\begin{equation} \label{rho_sol}
\rho^{2}\left(\xi\right)=\begin{cases}
\alpha_{d}\left(0\right)\left(\xi-\xi_{0}\right){}^{2}-\frac{\gamma_{d}\left(0\right)}{\alpha_{d}\left(0\right)} & m_{\text{eff}}^{2}=0,\alpha_{d}\left(0\right),\gamma_{d}\left(0\right)\ne0\\
2\sqrt{\gamma_{d}\left(0\right)}\left(\xi-\xi_{0}\right) & m_{\text{eff}}^{2}=0,\alpha_{d}\left(0\right)=0,\gamma_{d}\left(0\right)\ne0\\
\text{const} & m_{\text{eff}}^{2}=0,\alpha_{d}\left(0\right)=\gamma_{d}\left(0\right)=0\\
\frac{1}{2m_{\text{eff}}^{2}}\left[\alpha_{d}\left(m_{\text{eff}}^{2}\right)+\sqrt{\beta_{d}\left(m_{\text{eff}}^{2}\right)}\sin\left(2\sqrt{m_{\text{eff}}^{2}}\left(\xi-\xi_{0}\right)\right)\right] & m_{\text{eff}}^{2}>0\\
\frac{1}{2m_{\text{eff}}^{2}}\left[\alpha_{d}\left(m_{\text{eff}}^{2}\right)-\sqrt{-\beta_{d}\left(m_{\text{eff}}^{2}\right)}\sinh\left(2\sqrt{-m_{\text{eff}}^{2}}\left(\xi-\xi_{0}\right)\right)\right] & m_{\text{eff}}^{2}<0,\beta_{d}\left(m_{\text{eff}}^{2}\right)<0\\
\frac{1}{2m_{\text{eff}}^{2}}\left[\alpha_{d}\left(m_{\text{eff}}^{2}\right)-\sqrt{\beta_{d}\left(m_{\text{eff}}^{2}\right)}\cosh\left(2\sqrt{-m_{\text{eff}}^{2}}\left(\xi-\xi_{0}\right)\right)\right] & m_{\text{eff}}^{2}<0,\beta_{d}\left(m_{\text{eff}}^{2}\right)>0
\end{cases}
\end{equation}
where $\beta_d$ is given by:
\begin{equation}
    \beta_{d}\left(m_{\text{eff}}^{2}\right)=\alpha_{d}^{2}\left(m_{\text{eff}}^{2}\right)+4m_{\text{eff}}^{2}\gamma_d\left(m_{\text{eff}}^{2}\right).
\end{equation}
Furthermore, by integrating \eqref{Eptcl} and \eqref{Lptcl}, the temporal and angular coordinates can be expressed explicitly in terms of $\rho^2$:
\begin{multline*}
t=t_{0} \pm \frac{r_5}{2\left(\rho_{+}^{2}-\rho_{-}^{2}\right)}\left[\rho_{+}\sqrt{1+\frac{\rho_{+}^{2}}{R^{2}}}f^{-}\left(\rho^{2}-\rho_{+}^{2},B_{+},C_{+}\right)+\rho_{-}\sqrt{1+\frac{\rho_{-}^{2}}{R^{2}}}f^{^{+}}\left(\rho^{2}-\rho_{-}^{2},B_{-},C_{-}\right)\right] \\
\pm \frac{E}{R}\cdot h\left(\rho^{2}-\rho_{+}^{2},B_{+},C_{+}\right)
\end{multline*} 
\begin{multline}   \label{theta_sol}
\theta=\theta_{0}\pm\frac{r_5}{2\left(\rho_{+}^{2}-\rho_{-}^{2}\right)}\left[\rho_{-}\sqrt{1+\frac{\rho_{+}^{2}}{R^{2}}}f^{-}\left(\rho^{2}-\rho_{+}^{2},B_{+},C_{+}\right)-\rho_{+}\sqrt{1+\frac{\rho_{-}^{2}}{R^{2}}}f^{-}\left(\rho^{2}-\rho_{-}^{2},B_{-},C_{-}\right)\right] \\ 
\pm\frac{L}{R^{2}}\cdot h\left(\rho^{2}-\rho_{+}^{2},B_{+},C_{+}\right)
\end{multline}
with the auxiliary functions defined as:
\begin{align} 
f^{^{\pm}}\left(x,B,C\right) &= \log\left(\frac{\frac{x}{R^{2}}}{2C\sqrt{-m_{\text{eff}}^{2}x^{2}+Bx+C^{2}}\pm Bx\pm2C^{2}}\right)\\ h\left(x,B,C\right) &= \frac{r_{5}}{2\sqrt{m_{\text{eff}}^{2}}}\arctan\left(\frac{-2m_{\text{eff}}^{2}x+B}{2\sqrt{m_{\text{eff}}^{2}}\sqrt{-m_{\text{eff}}^{2}x^{2}+Bx+C^{2}}}\right)
\end{align}
and the coefficients given by:
\begin{equation} \label{B,C}
    B_{\pm}=E^{2}R^{2}-L^{2}\mp m_{\text{eff}}^{2}\left(\rho_{+}^{2}-\rho_{-}^{2}\right)+\frac{E^{2}R^{2}\left(\rho_{+}^{2}+\rho_{-}^{2}\right)-2ERL\rho_{+}\rho_{-}}{R^{2}},\quad C_{\pm}=\sqrt{1+\frac{\rho_{\pm}^{2}}{R^{2}}}\left(ER\rho_{\pm}-L\rho_{\mp}\right).
\end{equation}
To find the geodesics for the canonically normalized radial and angular coordinates $\phi,x$ one can directly apply the transformations \eqref{canon_norm} to the solutions \eqref{rho_sol}-\eqref{theta_sol}.

The functional form of the radial solution looks similar to the one obtained for a pure BTZ black hole background \cite{Cruz:1994ir}.
In the pure BTZ case, timelike geodesics of probe particles can starting from the past horizon reach maximal finite radius $\rho_{max}>\rho_+$ before falling back towards the singularity if they possess small angular momentum. For sufficiently large angular momentum, they exhibit oscillatory motion between this maximum radius and a non-zero minimal radius $\rho_{min}$ located below the inner horizon. Meanwhile, spacelike trajectories can reach the boundary $\rho\to \infty$ in a finite proper time. 
For null trajectories,  massless particles reach a maximum radius at the boundary $\rho_{max}\to\infty$, where they are reflected and subsequently fall back toward the black hole. In the extremal black hole background , particles can orbit the black hole in a circular motion at any radius for specific values of their conserved quantities; stable circular orbits also appear for timelike trajectories precisely on the outer horizon \cite{Farina:1993xw}.

In our asymptotically flat background, this qualitative picture changes significantly due to the introduction of the effective mass squared in \eqref{meff}. For the $\lambda$-deformed BTZ background, timelike trajectories can escape to the asymptotically flat region, and reach the timelike infinity, provided that their positive mass squared lies \textit{below} a certain critical non-negative threshold $m_c^2$, which is determined by the particle's energy $E$, angular momentum $L$ and the asymptotic radius $R$:
\begin{equation} \label{massthreshold}
    0\le m^2\le m_c^2 = \frac{E^{2}R^{2}-L^{2}}{R^{2}},\qquad ER\ge L
\end{equation}
Heavier particles fall into the regime where $m_{\text{eff}}^2>0$, and are bounded, reaching a finite maximum radius at:
\begin{equation} \label{rhomax}
 \rho_{max}^2 = \frac{\alpha_d(m_{\text{eff}}^2)+\sqrt{\beta_d(m_{\text{eff}}^2)}}{2m_{\text{eff}}^2}.
\end{equation}
These particles are safe from hitting the singularity if $\alpha_d(m_{\text{eff}^2})>\sqrt{\beta_d(m_{\text{eff}^2})}$, instead reaching a minimal radius $0<\rho_{min}<\rho_-$.

When $\alpha_d(0)=\gamma_d(0)=0$ the trajectory corresponds to circular motion at an arbitrary radius $\rho$.
Tachyonic particles following spacelike trajectories $m^2<0$ always reach spacelike infinity whenever $ER\ge L$. Furthermore, if the critical mass squared is negative ($m_c^2<0$ which occurs when $L>ER$) they will still escape to spacelike infinity as long as $m^2 \le m_c^2$. Otherwise, similar to  timelike trajectories in pure BTZ case, "light" tachyons and heavier particles are bounded, reaching a finite maximum radius $\rho=\rho_{max}$ and then be falling back.

\subsection{Particle Emission from the Deformed BTZ}
In this section, we will compute the probability for a particle carrying energy $E$ and angular momentum $L$ to be emitted from the vicinity in the outer horizon $\rho=\rho_+$ of a deformed BTZ black hole with a energy $E_{BH}$ and angular momentum $J_{BH}$. The emission process leaves behind a remnant black hole with energys $E_{BH}-E$ and angular momentum $J_{BH}-L$.
This calculation is preformed in the  Gullstrand–Painlevé coordinate system, which is suited for analyzing near-horizon physics because it eliminates the coordinate singularity that appears at the outer horizon in the original coordinate system \eqref{defbtz}.
The explicit coordinate transformation is given by\footnote{Notice that we the canonical normalization of the time coordinate $t$ throughout our analysis.}:

\begin{equation} \label{PG}
  dt=d\mathfrak{t}-\frac{R}{r_{5}}\sqrt{\frac{1}{N^{2}}\left(\frac{1}{N^{2}}-\frac{1}{f(\rho)}\right)\left(1+\frac{\rho^{2}}{R^{2}}\right)}d\rho,\quad d\theta=d\varphi-N_{\theta}\sqrt{\frac{1}{N^{2}}\left(\frac{1}{N^{2}}-\frac{1}{f(\rho)}\right)\left(1+\frac{\rho^{2}}{R^{2}}\right)}d\rho
\end{equation}
where $f(\rho)$ is a well defined function for all $\rho$\footnote{For simplicity, one can choose $f(\rho)=1$ as in \cite{Wu:2006pz} for the standard BTZ black hole. Other choices include $f(\rho)=\rho^2+\frac{\rho_-\rho_+}{\rho^2}$, which take into account angular potential in \cite{Wu:2006pz}, or  $f(\rho)=1+\rho^2$, hich compatible with global AdS space as shown in \cite{Hemming:2000as} }. The metric, in Gullstrand–Painlevé coordinates $(\mathfrak{t},\rho,\varphi)$ is:
\begin{equation} \label{G-P}   ds^{2}=\frac{d\rho^{2}}{f(\rho)}+\frac{1}{1+\frac{\rho^{2}}{R^{2}}}\left(-\frac{r_{5}^{2}}{R^{2}}N^{2}d\mathfrak{t}^{2}+\frac{2r_{5}}{R}\sqrt{N^{2}\left(\frac{1}{N^{2}}-\frac{1}{f(\rho)}\right)\left(1+\frac{\rho^{2}}{R^{2}}\right)}d\rho d\mathfrak{t}+\rho^{2}\left(d\varphi-\frac{r_{5}}{R}N_{\theta}d\mathfrak{t}\right)^{2}\right)
\end{equation}

We aim to evaluate the emission probability amplitude, which in the tunneling approach corresponds to the transmission coefficient of a radial trajectory across the classically forbidden region of the deformed BTZ horizon barrier. Within this approach,
the tunneling rate is governed by the imaginary part of the particle's Hamiltonian action, integrated along a radial trajectory from the initial horizon $\rho_{in}=\rho_+(E_{BH},J_{BH})$ to the new horizon $\rho_{out}=\rho_+(E_{BH}-E,J_{BH}-L)$ that established itself after the emission.
The probability of quantum to escape the black hole is proportional to:
\begin{equation} \label{Gamma1}
    \Gamma \sim \exp\left(-2\Im \mathcal{I}\right)
\end{equation}
where $\mathcal{I}$ represents the Hamiltonian action for the trajectory. For  particle emission, this action is given by the Legendre transform of the worldline action \eqref{worldline}:
\begin{equation} \label{I_ptcl}
   \mathcal{I}_{\text{ptcl}}=\int d\xi\left(p_{\mu}\dot{x}^{\mu}-n_0\mathcal{H}_{ptcl}\right) 
\end{equation}
where $\mathcal{H}$ is the Hamiltonian density \eqref{H_constr} and $n_0=\frac{1}{2r_5}$ is the Lagrange multiplier. Evaluating the path integral over the classically forbidden region yields:
 \begin{equation} \label{H_path}
 \Im\mathcal{I}_{\text{ptcl}}=\Im\left(\int_{\rho_{in}}^{\rho_{out}}d\rho\,p_{\rho}\right)=\Im\left(\int_{\rho_{in}}^{\rho_{out}}d\rho\,\int_{0}^{p_{\rho}}dp_{\rho}'\right)  
 \end{equation}
The second relation encapsulates the core assumption for the tunneling method \cite{Parikh:1999mf}, in which the particles radial momentum is modified due to the dynamical adjustment of the background geometry during the emission process, represented by the identification $p_\rho\to dp_\rho$. 
We note that any potential contributions arising from angular and temporal trajectories  integrated over their conjugate momenta $p_\frak{t},p_{\varphi}$ are purely real, and therefore omitted from the calculation of the imaginary part of the action.
To evaluate this integral, we compute the radial momentum to leading order in $\epsilon = \rho-\rho_+$ in terms of the emitted particle energy $E$ and angular momentum $L$. Utilizing the geodesic constraints \eqref{Eptcl}-\eqref{H_constr}\footnote{In this context, $E,L$ are the conjugate momentum of the $\frak{t},\varphi$ coordinates, respectively, and the metric $G_{\mu\nu}$ entering the Hamiltonian constraint \eqref{H_constr} is replaced by the Gullstrand–Painlevé metric \eqref{G-P}.} we choose the outgoing mode:
\begin{equation} \label{pr}
    p_{\rho}  \sim \frac{r_{5}\sqrt{1+\frac{\rho_{+}^{2}}{R^{2}}}\left(ER\rho_{+}-L\rho_{-}\right)}{\left(\rho-\rho_{+}\right)\left(\rho_{+}^{2}-\rho_{-}^{2}\right)}
\end{equation}
We observe that the leading-order radial momentum on the horizon is independent of the particle's mass squared $m^2$ , reflecting the fact that the horizon crossing is locally a null trajectory.

By varying the black hole parameters, the differential change in the radial momentum near the horizon can be expressed via the variations of the background geometry. From \eqref{pr}, we find:
\begin{equation} \label{dpr_partic}
    dp_{\rho} = -\frac{Rr_{5}\rho_{+}\sqrt{1+\frac{\rho_{+}^{2}}{R^{2}}}}{\left(\rho-\rho_{+}\right)\left(\rho_{+}^{2}-\rho_{-}^{2}\right)}\left(d\tilde{E}_{BH}-\Omega_{BH}d\tilde{J}_{BH}\right) 
\end{equation}
where $\Omega_{BH}$ is the angular velocity of the deformed BTZ  \eqref{S,T,OM}, and we introduce the intermediate quantities $\tilde{E}_{BH} =E_{BH}-E$ and $\tilde{J}_{BH}=J_{BH}-L$.
Substituting this differential momentum back into the path integral and reversing the order of integration to perform the radial coordinate integral first, we obtain: \begin{equation}
\Im\mathcal{I}_{\text{ptcl}}=-\Im\left(\,\int_{\left(E_{BH},J_{BH}\right)}^{\left(E_{BH}-E,J_{BH}-L\right)}\frac{Rr_{5}\rho_{+}\sqrt{1+\frac{\rho_{+}^{2}}{R^{2}}}}{\left(\rho_{+}^{2}-\rho_{-}^{2}\right)}\left(d\tilde{E}_{BH}-\Omega_{BH}d\tilde{J}_{BH}\right)\int_{\rho_{in}}^{\rho_{out}}\frac{d\rho}{\rho-\rho_{+}}\right)
\end{equation}
Applying Sokhotski–Plemelj theorem by introducing the standard analytical continuation $\rho_+\to\rho_{+}-i\varepsilon$ and taking the limit $\varepsilon\to 0$, the complex integral decomposes into its real and imaginary parts:

\begin{equation} \label{S-P}
  \int^{\rho_{out}}_{\rho_{in}} \frac{d\rho}{\rho-\rho_+-i\varepsilon} =\mathcal{P} \int^{\rho_{out}}_{\rho_{in}} \frac{d\rho}{\rho-\rho_+}+i\pi,\qquad \mathcal{P} \int^{\rho_{out}}_{\rho_{in}} \frac{d\rho}{\rho-\rho_+}\in\mathbb{R} 
\end{equation}
where $\mathcal{P}$ denotes the Cauchy principle value over the real line.
Extracting the  imaginary component yields the final expression for the action:
\begin{equation} \label{ImI}
\Im\mathcal{I}_{\text{ptcl}}=-\int_{\left(E_{BH},J_{BH}\right)}^{\left(E_{BH}-E,J_{BH}-L\right)}\frac{1}{2T_{BH}}\left(d\tilde{E}_{BH}-\Omega_{BH}d\tilde{J}_{BH}\right)=-\frac{1}{2}\int_{S\left(E_{BH},J_{BH}\right)}^{S\left(E_{BH}-E,J_{BH}-L\right)}dS_{BH}.
\end{equation}
Here, $T_{BH}$ is the deformed BTZ temperature given in \eqref{S,T,OM}. In the second equality, we have implemented the first law of thermodynamics \eqref{First Law}, treating the black hole system as evolving within the microcanonical ensemble during the particle quanta emission.
Combining \eqref{Gamma1} and \eqref{ImI}, the total emission probability for the particle quantum reduces directly to the change in the Bekenstein-Hawking entropy of the background:
\begin{equation} \label{Gamma2}
    \Gamma \sim \exp\left(\Delta S_{BH}\right)
\end{equation}
This result matches the behavior established for the standard BTZ black hole \cite{Wu:2006pz,Li:2006rg}, as well as early foundational studies of particle tunneling from Schwarzschild-like geometries \cite{Parikh:1999mf,Vanzo:2011wq}. 
Here, it is generalized to a family of backgrounds with asymptotically flat spacetimes and linear dilaton \eqref{defbtz}-\eqref{dilaton}.
As discussed in \cite{Martinec:2023plo}, an alternative way to to obtain this result is through the WKB approach. In this approach, the imaginary part of the Hamiltonian path integral action is calculated on a fixed background, without explicitly factoring in backreaction effects at the leading order; backreaction is instead treated as a higher-order perturbation.

In this approach, the energy and angular momentum of the emitted particle are directly identified with the differentials of the black hole charges, transforming \eqref{H_path} into:
\begin{equation}
    \Im\mathcal{I}_{\text{ptcl}}=\Im\left(\int_{\rho_{in}}^{\rho_{out}}d\rho\,p_{\rho}\right)=\Im\left(\int_{\rho_{in}}^{\rho_{out}}d\rho   \frac{r_{5}\sqrt{1+\frac{\rho_{+}^{2}}{R^{2}}}\left(ER\rho_{+}-L\rho_{-}\right)}{\left(\rho-\rho_{+}\right)\left(\rho_{+}^{2}-\rho_{-}^{2}\right)}\right).  
\end{equation}
Performing the contour integration by analytical continuation around the horizon pole into the lower half complex plane, in complete analogy with \eqref{S-P}, leads to:
\begin{equation} \label{WKB_part}
   \Im\mathcal{I}_{\text{ptcl}}= \frac{1}{2T}\left(E-\Omega L\right)=-\frac{1}{2T_{BH}}(d\tilde{E}_{BH}-\Omega_{BH}d\tilde{J}_{BH}) = -\frac{1}{2}dS_{BH}
\end{equation}
where we associate $E=-d\tilde{E}_{BH},L=-d\tilde{J}_{BH}$ matching the background parameters $\Omega=\Omega_{BH}$ and $T=T_{BH}$ which  reproduces the universal result \eqref{Gamma2}.
One can further cross-check this result by directly computing the finite difference in the entropy using its explicit Cardy form:
\begin{equation}
    \Delta S_{BH}=S_{BH}(E_{BH}-E,J_{BH}-L)-S_{BH}(E_{BH},J_{BH})
\end{equation}
where the entropy formula \eqref{Entropy} for an arbitrary state with energy $e$ and angular momentum $\ell$ is:
\begin{equation}
    S_{BH}(e,j)= 2\pi \sqrt{kp}\left(\sqrt{e_{L}\left(1+\frac{\lambda}{p}e_{R}\right)}+\sqrt{e_{R}\left(1+\frac{\lambda}{p}e_{L}\right)}\right),\qquad e_{L,R}(\lambda)=\frac{1}{2}\left(eR\pm \ell\right). 
\end{equation}
Expanding $\Delta S_{BH}$ 
to linear order in $\frac{E}{E_{BH}}\ll 1 $ and $\frac{L}{J_{BH}}\ll 1$, and substituting the explicit definitions for $E_{BH}$ and $J_{BH}$ from \eqref{M,J}-\eqref{E_BH+},
yields precisely the  linear relationship found in \eqref{WKB_part} in terms of the conserved quantities $E,L$.
A direct physical consequence of \eqref{Gamma2} is that low-energy particles are much more likely radiated than high-energy ones; high-energy emissions cause a large geometric shift that significantly reduces the remaining entropy, which exponentially suppresses the process.

In the next subsections, we generalize this point-particle framework to the case of a winding long string, demonstrating  both of these semiclassical paradigms.

\subsection{Winding Long String Geodesics}
\label{sec:pos_string}
The geodesics of a probe long $w$-winding string can be obtained via the \textit{spectral flow operation} \cite{Maldacena:2000hw,Hemming:2001we}, such that the unwound $w=0$ states correspond to the particle geodesics treated in the previous section \eqref{rho_sol}-\eqref{theta_sol}.
The spectral flow operation translates into the following coordinates transformations:
\begin{equation} \label{spectralflow}
    t_{string}(\xi,\sigma)=t(\xi)+w\xi,\qquad \theta_{string}(\xi,
    \sigma) =\theta(\xi)+w\sigma, \qquad \rho_{string}(\xi,\sigma) = \rho(\xi)
\end{equation}
The spectral flow operation was originally applied  to the
standard BTZ case to generate a continuous representation of long string states with an unbounded energy spectrum. Here, we provide a qualitative argument for why these same spectral flow rules \eqref{spectralflow}  also applied to long strings on the $\lambda$-deformed BTZ background \eqref{defbtz}-\eqref{dilaton}:

In the deep IR limit $\frac{\rho}{R}\ll 1$, the geometry reduces to the standard, undeformed BTZ background \eqref{btxmetric}.
String theory on the BTZ background can be realized as an exact worldsheet theory, given as a WZW model based on the $SL(2,\mathbb{R})_k$ group, 
having the action:
\begin{equation} \label{WZW}
    I_{\text{WZW}}[g]=\frac{k}{4\pi}\left[\int_{\Sigma}\text{Tr}\left(g^{-1}\partial gg^{-1}\bar{\partial}g\right)-\frac{1}{3}\int_{D}\text{Tr}\left(g^{-1}dg\right)^{3}\right]
\end{equation}  
Here, the worldsheet manifold $\Sigma$  constitutes the boundary of the 3-dimensional manifold (i.e. $\Sigma=\partial D$), and the group element $g\in SL(2,\mathbb{R})$ is parametrized as:
\begin{equation} \label{gBTZ}
    g=e^{\frac{\rho_{+}-\rho_{-}}{2r_{5}}\left(\frac{\tau}{r_{5}}+\theta\right)\sigma_{3}}e^{r\sigma_{1}}e^{-\frac{\rho_{+}+\rho_{-}}{2r_{5}}\left(\frac{\tau}{r_{5}}-\theta\right)\sigma_{3}}
\end{equation} 
where $\{\sigma_i\}_{i=1}^{3}$ are the Pauli matrices, and $r$ is given in terms of the radial $\rho$ coordinate and the horizons $\rho_{\pm}$ such that:
\begin{equation} \label{redf}
\cosh^{2}\left(r\right)=\frac{\rho^{2}-\rho_{-}^{2}}{\rho_{+}^{2}-\rho_{-}^{2}}.
\end{equation}
To properly describe the BTZ background, the group element $g$ must be accompanied by a discrete group identification:
\begin{equation}
    g\sim e^{\frac{\pi(\rho_+-\rho_-)}{r_5}\sigma_3}ge^{\frac{\pi(\rho_++\rho_-)}{r_5}\sigma_3}
\end{equation}
which is equivalent to imposing a periodic condition on the angular coordinate $\theta\sim \theta + 2\pi$.
The spectral flow transformations \eqref{spectralflow} can be realized as a global $U(1)_L \times U(1)_R$ automorphism gauge transformation on the group manifold, such that:
\begin{equation}
g\to e^{\frac{(\rho_{+}-\rho_{-})}{r_{5}}w^{+} z\sigma_{3}}ge^{-\frac{(\rho_{+}+\rho_{-})}{r_{5}}w^{-} \bar{z}\sigma_{3}},\quad (z,\bar{z})=\frac{1}{r_{5}}\xi\pm\sigma
\end{equation}
The identification of the left and right -moving winding numbers $w^{-}=w^{+}=w$ ensures there is no closed timelike curve on $SL(2,
\mathbb{R})$.

Conversely, in the asymptotic UV limit of the deformed background $\frac{\rho}{R}\gg1$, the geometry approaches a flat spacetime accompanied by a linear dilaton and a constant B-field  \eqref{Fmetric}-\eqref{FBdialaton}. String theory in flat spacetime exhibits a trivial winding symmetry in the $t-\theta$ sector. In this asymptotic regime, the worldsheet equations of motion for $t$ and $\theta$ reduce to free wave equations, which are trivially invariant under linear shifts of the worldsheet coordinates. For a closed long string, the worldsheet spatial coordinate is periodic ($\sigma \sim \sigma + 2\pi$). Long strings describe scattering states that have a support in the flat asymptotic region and carry a winding number $w$ around the compact coordinate; thus, they must satisfy the topological boundary condition $\theta(\xi, \sigma+2\pi) = \theta(\xi, \sigma) + 2\pi w$. This requirement guarantees that the exact linear shifts defined by the spectral flow in \eqref{spectralflow} represent valid classical trajectories, ensuring the symmetry holds in this limit.

We therefore find that the spectral flow \eqref{spectralflow} represents a  symmetry at both asymptotic extremes of the spacetime: the purely curved 
$SL(2,\mathbb{R})$
 WZW model in the IR, and the free compact boson CFT in the UV. Because the target space isometries associated with translations in 
$t,\theta$ remain globally unbroken throughout the interpolating bulk geometry, the kinematic coordinate maps defined by the standard spectral flow must remain valid classical trajectories for winding long strings propagating anywhere within the full $\lambda$-deformed BTZ background\footnote{An indication for this assumption also appears in the worldsheet analysis for string theory on TsT black holes in \cite{Apolo:2019zai}.}. 

As in the pure BTZ case, the deformed theory can be realized  an exact worldsheet action, given by the coset $\frac{SL(2,\mathbb{R})_k\times U(1)}{U(1)}$ gauged-WZW model \cite{Chakraborty:2020yka,Chakraborty:2022dgm}.
In Appendix \ref{appB}, we review the details of the derivation of the coset worldsheet action and show precisely how to obtain the $\lambda$-deformed BTZ from an underlying BTZ group manifold \eqref{gBTZ}.

It is worth mentioning that the worldsheet structure of a similar family of deformed solutions, obtained by solution-generating transformations, a.k.a TsT-black holes \cite{Apolo:2019zai} (see also \cite{Araujo:2018rho}), allows one to describe long string states propagating in the deformed background by appealing to an equivalent picture.
In this dual picture,  the long string propagates in the undeformed background, accompanied by a modified state-dependent spectral flow given in terms of the excitation energy $E_{w,string}$ and the angular momentum $L_{string}$ of the long string.
It can be shown that the  $\lambda$-deformed BTZ solution \eqref{defbtz}-\eqref{dilaton} can also be obtained via TsT transformations, giving rise to a background-dependent spectral flow in addition to a state-dependent one. 
However, the generated solution is highly sensitive to the value of the B-field at the origin $B_{\tau\theta}^{(0)}$, of the original undeformed solution \cite{Apolo:2025wcl}. To obtain the $\lambda$-deformed BTZ solution, the TsT approach dictates a $B_{\tau\theta}^{(0)}$ value that is non-trivial and  deformation-dependent. This is in tension with previous results established in \cite{Martinec:2023plo,Chakraborty:2024ugc,Giveon:2024sgz}.

The total energy and angular momentum of the long string in the deformed BTZ background are obtained by evaluating the on-shell condition for the exact vertex operator in the asymptotically flat regime $\rho\to\infty$ where long strings dominate, as described in \cite{Giveon:2024sgz} (and for the non-rotating case in \cite{Chakraborty:2024ugc}). 
The spectrum of the $w$-excited long string coincide with the $\mathbb{Z}_w$ twisted sector of a single trace $T\bar{T}$ deformed symmetric orbifold $\mathcal{M}^p/S_p$:
\begin{equation} \label{TTform}
 E_{w,total}(\lambda)=\frac{w}{\lambda R}\left(-1+\sqrt{1+\frac{2\lambda \mathcal{E}_{w,total}}{w}+\left(\frac{\lambda L_{string}}{w}\right)^2}\right),\quad \lambda = \frac{\alpha'}{R^2}>0
\end{equation}
where $E_{w,total}$ is the total energy of the emitted string. This equals the deformed excitation energy of the $w$-winding long string, $E_{w,string}$, plus the the contribution of $w$ out of $p$ strings to the energy of the deformed BTZ background, while the remaining $p-w$ winding string contribute to the residual part of the deformed BTZ energy $E_{BH}$:
\begin{equation} \label{total deformed}
E_{w,total}R=E_{w,string}R+w\frac{E_{BH}R}{p} 
\end{equation}
It was shown in \eqref{pos_TT_energy} that black holes energy $E_{BH}$ coincides with the maximally twisted $\mathbb{Z}_p$ sector of a single trace $T\bar{T}$ deformed symmetric orbifold $\mathcal{M}^p/S_p$ thus said to be in harmony with single trace $T\bar{T}$ holography. Equation \eqref{total deformed} generalizes the idea further to include energies of long string propagating in the deformed background.

Here, $\mathcal{E}_{w,total}$ represents the dimensionless total energy of the excited string above the undeformed BTZ black hole background, which includes the undeformed excitation energy plus contribution to the BTZ black hole's mass $M_{BTZ}$:
\begin{equation} \label{undef tot}
\mathcal{E}_{w,total}=\mathcal{E}_{w,string}+w\frac{r_5M_{BTZ}}{p}
\end{equation}
such that\footnote{In particular, the same holds for the excitation energy: $\lim_{R/\sqrt{\alpha'}\to\infty}E_{w,string}R = \mathcal{E}_{w,string}$ }:
\begin{equation} \label{lim_undef}
 \lim_{R/\sqrt{\alpha'}\to \infty}E_{w,total}R = \mathcal{E}_{w,total}.
\end{equation}
The quantities $\mathcal{E}_{w,total}$ and $L_{string}$ are associated with the  worldsheet CFT $N_{L,R}$ oscillatory levels, subject to the full string theory BRST constraints, and the continuous unbounded radial momentum of the long string $\frac{2s}{r_5}$, which is represented by the quantum number $j=-\frac{1}{2} +is$ with $s\in\mathbb{R} $:

\begin{equation}
\begin{aligned}
& \mathcal{E}_{w,total} = \frac{1}{w}\left(-\frac{2j(j+1)}{k}+N_L+N_R -1\right) \\
& L_{string} = \frac{1}{w}\left(N_L-N_R\right)\in \mathbb{Z}
\end{aligned}
\end{equation}
We interpret these definitions in the context of a process where we extract fundamental string charge $w$ to make a smaller black hole, such that $\delta{E_{BH}}=-\frac{w}{p}E_{BH}$, alongside an emitted string that carries winding $w$, energy and angular momentum above the threshold $E_{w,total}$ and $L_{string}$.

In \cite{Giveon:2024sgz}, it was found that the formula in \eqref{total deformed}  which ensures harmony with single trace $T\bar{T}$ holography for the long string spectrum, holds if the asymptotic B-field value is chosen to be a specific critical value. For a given solution \eqref{btautheta}, this firmly fixes the B-field at the origin to be:
\begin{equation} \label{B^0_p}
 B_{tx}^{(0)} = -\frac{\frac{\rho_-^2\rho_+^2}{R^4}}{\sqrt{\left(1+\frac{\rho_-^2}{R^2}\right)\left(1+\frac{\rho_+^2}{R^2}\right)}} \Rightarrow B_{\tau \theta}^{(0)} =-\frac{R^2}{r_5}\frac{\frac{\rho_-^2\rho_+^2}{R^4}}{\sqrt{\left(1+\frac{\rho_-^2}{R^2}\right)\left(1+\frac{\rho_+^2}{R^2}\right)}} 
\end{equation}
In the next section, we will provide an independent thermodynamic justification for this particular choice, deriving it from its  affect on black hole thermodynamics, where the B-field at the horizon acts as the chemical potential for the emission of winding strings.

Long strings in the deformed background exhibit bounded trajectories when derived from a heavy particles following a timelike trajectories (when $ER>L$), and include also spacelike and null trajectories (when $ER<L$), as can be read 
from the spectral flow operation \eqref{spectralflow}.
Unbounded long string trajectories arise  from spacelike particle trajectories, and this extended to include timelike particles possessing a state-dependent mass squared threshold \eqref{massthreshold} in the case where $ER>L$.

Each point on the string corresponds to a timelike trajectory that reaches timelike infinity $\rho,t\to \infty$. The long string propagates in the radial direction with an additional rate $s$ (associated with the quantum number $j=-\frac{1}{2}+is$), and in the temporal direction at rate $|w|$ relative to that of a particle.
The arrow of time dictated by the overall sign of the time rate, while the radial direction (inwards or outwards) dictated by the overall sign of the radial rate.

\subsection{Winding Long String Emission from the Deformed BTZ}
\label{2.4}
The process of emitting a $w$-winding long string from a deformed BTZ black hole, which initially possesses $p$ fundamental string charge, modifies the first law of thermodynamics \eqref{First Law}. Because the winding charge is now allowed to fluctuate, the system is described by the grand canonical ensemble:
\begin{equation} \label{grand_canon_1}
    dS_{BH} = \frac{1}{T_{BH}}\left(dE_{BH}-\Omega_{BH}dJ_{BH}-\mu_{BH}dp\right).
\end{equation}
The associated chemical potential is computed by contracting the  Killing horizon vector $\zeta^a_{H}$ with the azimuthal projection of the B-field evaluated at the horizon \cite{Compere:2007vx}:
\begin{equation}
    \mu_{BH}=-\frac{1}{\alpha'}\zeta_{H}^{\alpha} B_{\alpha\theta}|_{\rho=\rho_+},\qquad \zeta_H^\alpha=\partial_t-\Omega_{BH}\partial_\theta.
\end{equation}
This potential is highly sensitive to the choice of the B-field at the origin, $B_{t\theta}^{(0)}$:
\begin{equation} \label{chemical_pos}
    \mu_{BH}=-\frac{1}{\alpha'}\left(\frac{\rho_{+}^{2}}{R}\sqrt{\frac{1+\frac{\rho_{-}^{2}}{R^{2}}}{1+\frac{\rho_{+}^{2}}{R^{2}}}}+B_{t\theta}^{\left(0\right)}\right).
\end{equation}
To compute the tunneling process of the $w$-winding string from the vicinity of the black hole, we again choose to work in the regular Gullstrand–Painlevé coordinates \eqref{PG}. For the long strings, the B-field enters to the Hamiltonian constraint, meaning potential effects from new components $B_{\mathfrak{t}\rho},B_{\varphi\rho}$ must be considered. However, leveraging the gauge invariance of the B-field $B_{\mu\nu}\to B_{\mu\nu} +\partial_\mu C_\nu- \partial_\nu C_\mu$ for arbitrary vector field $C_{\mu}$, one can select a suitable gauge to make these specific components vanish. Such a choice included, for example:
\begin{equation} \label{gaugefix}
   C_{\rho} =\mathfrak{t}B_{\rho\frak{\mathfrak{t}}}\left(\rho\right)-\varphi B_{\varphi\rho}\left(\rho\right) ,\qquad C_{\frak{t}},C_{\varphi}=const. 
\end{equation}
This transformation safely removes the radial cross-terms while leaving the critical temporal-angular component unchanged, i.e.,  $B_{t\theta}=B_{\frak{t}\varphi}$. The emitted $w$-winding long string will therefore be governed by a Hamiltonian path integral given by:
\begin{equation} \label{Istr}
    \mathcal{I}_{string}=\int (p_{\mu}\dot{x}^\mu-n_0\mathcal{H}_{string}-n_1 P_{string})
\end{equation}
where the string Hamiltonian constraint is modified accordingly:
\begin{equation} \label{H_st_con}
    \mathcal{H}_{string} = G^{\mu\nu}\Pi_{\nu}\Pi_{\mu}+\frac{1}{\alpha'^{2}}G_{\mu\nu}x'^{\mu}x'^{\nu}+m^{2}=0
\end{equation}
An additional momentum constraint reflects the spatial  diffeomorphism invariance along worldsheet $\sigma$ coordinate:
\begin{equation} \label{P_constr}
P_{string} = p_{\mu}x'^{\mu} = 0
\end{equation}
Here, $\Pi_{\mu}$ is the kinetic momentum, which encodes the affect of the B-field on the geodesics:
\begin{equation} \label{KinPI}
    \Pi_{\mu}=p_{\mu}-\frac{1}{\alpha'}B_{\mu\nu}x'^{\nu}.
\end{equation}
The effect of the dilaton field $\Phi(\rho)$ on the geodesics is suppressed due to the Weyl invariance of the worldsheet action \eqref{I_WS}. The flat worldsheet metric  \eqref{WS_metric} ensures that $\Phi(\rho)R^{(2)}$ appears strictly as topological term, which can be integrated out from the Hamiltonian path integral\footnote{This is manifested in particular by the identification $R^{(2)}=0$. In Appendix \ref{appA}, we will show that for general worldsheet metric, the dilaton field has a direct, non-trivial affect on the string dynamics.}.

Solving the Hamiltonian constraint \eqref{H_st_con} by using the spectral flow transformations \eqref{spectralflow}:
\begin{equation} \label{momentums}
    p_{\mathfrak{t}}=-E_{w,string},\quad p_{\varphi}=L_{string};\qquad \varphi'=w,\quad \rho'=\mathfrak{t}'=0.
\end{equation}
One can then explicitly find the radial momentum, $p_{\rho}$, which in the near horizon limit at leading order, is given by: 
\begin{equation}
    p_{\rho}\sim\frac{r_{5}\sqrt{1+\frac{\rho_{+}^{2}}{R^{2}}}}{\left(\rho-\rho_{+}\right)\left(\rho_{+}^{2}-\rho_{-}^{2}\right)}\left(E_{w,string}R\rho_{+}-L_{string}\rho_{-}+\frac{w R\rho_{+}}{\alpha'}\left(\frac{\rho_{+}^{2}}{R}\sqrt{\frac{1+\frac{\rho_{-}^{2}}{R^{2}}}{1+\frac{\rho_{+}^{2}}{R^{2}}}}+B_{t\theta}^{\left(0\right)}\right)\right)
\end{equation}
Following the same steps utilized for the particle emission rate in the tunneling approach \eqref{H_path}-\eqref{ImI},
 we now make the thermodynamic identifications
$d E_{w,string}=-d\tilde{E}_{BH},\quad L_{string}=-d\tilde{J}_{BH}$, and $w=-d\tilde{p}$, where:
\begin{equation}
    \tilde{E}_{BH} =E_{BH}-E_{w,string},\quad \tilde{J}_{BH}=J_{BH}-L_{string},\quad \tilde{p}=p-w
\end{equation}
Integrating this complex path yields:

\begin{equation} \label{ImIst}
\Im\mathcal{I}_{\text{string}}=-\int_{\left(E_{BH},J_{BH},p\right)}^{\left(E_{BH}-E,J_{BH}-L,p-w\right)}\frac{1}{2T_{BH}}\left(d\tilde{E}_{BH}-\Omega_{BH}d\tilde{J}_{BH}-\mu_{BH}d\tilde{p}\right)=-\frac{\Delta S_{BH}}{2}.
\end{equation}
with the potentials  $T_{BH},\Omega_{BH}$  given by \eqref{S,T,OM}, and the chemical potential given by \eqref{chemical_pos}.
The result in \eqref{ImIst} recovers the same universal entropy law \eqref{Gamma2}
that was derived for the particle emission in the microcanonical ensemble.

As in the particle case, this result can also be obtained via the WKB approach by  applying the differentials $E_{w,string}=-d\tilde{E}_{BH},L_{w,string}=-d\tilde{J}_{BH}$, and $w=-dp$.
Using the Cardy form for the entropy \eqref{Entropy}, we can express the exact difference between the initial black hole entropy and that of the remnant black hole after the emission of a long string. This emitted string carries a total energy  $E_{w,total}$ \eqref{total deformed}, angular momentum $L_{string}$, and winding charge $w$ out of $p$ from the original setup:
\begin{equation} \label{DeltaS}
 \Delta S_{BH}=S_{BH}(E_{BH}-E_{w,total},J_{BH}-L_{string},p-w)-S_{BH}(E_{BH},J_{BH},p)
\end{equation}
Explicitly, this expands to:
\begin{multline}
 \Delta S_{BH}=2\pi \sqrt{k(p-w)}\left(\sqrt{E^{\Delta}_{L}\left(1+\frac{\lambda}{p-w}E^{\Delta}_{R}\right)}+\sqrt{E^{\Delta}_{R}\left(1+\frac{\lambda}{p-w}E^{\Delta}_{L}\right)}\right)-\\ 2\pi \sqrt{kp}\left(\sqrt{E_{L}\left(1+\frac{\lambda}{p}E_{R}\right)}+\sqrt{E_{R}\left(1+\frac{\lambda}{p}E_{L}\right)}\right)
\end{multline}

where $E_{L,R}$ are given in \eqref{ELR}, and the remnant energies are:
\begin{equation}
E^{\Delta}_{L,R} = \frac{1}{2}\left[\left(E_{BH}-E_{w,total}\right)R\pm \left(J_{BH}-L_{string}\right)\right]
\end{equation}
Expanding \eqref{DeltaS} to linear order in $\frac{E_{w,string}}{E_{BH}},\frac{L_{string}}{J_{BH}}\ll 1$, and now $\frac{w}{p} \ll 1$, yields:
\begin{equation} \label{dS_BH_sol}
dS_{BH}=-\frac{2\pi r_{5}\sqrt{1+\frac{\rho_{+}^{2}}{R^{2}}}}{\rho_{+}^{2}-\rho_{-}^{2}}\left(E_{w,string}R\rho_{+}-L_{string}\rho_{-}+\frac{w}{\alpha'}\frac{\rho_{+}^{3}}{\sqrt{\left(1+\frac{\rho_{+}^{2}}{R^{2}}\right)\left(1+\frac{\rho_{-}^{2}}{R^{2}}\right)}}\right).
\end{equation}
By rearranging this expression and identifying the emitting string parameters as the differential changes in the thermodynamic quantities of the deformed BTZ black hole according to the WKB prescription, one finds that the chemical potential will have a unique, particular value:
\begin{equation} \label{pref_chem_pos}
 \mu_{BH}=-\frac{1}{\alpha'}\frac{\rho_{+}^{2}}{R\sqrt{\left(1+\frac{\rho_{+}^{2}}{R^{2}}\right)\left(1+\frac{\rho_{-}^{2}}{R^{2}}\right)}}
\end{equation}
Comparing this required thermodynamic potential to \eqref{chemical_pos}, reveals that there is a uniquely favored value for the B-field at the origin:
\begin{equation} \label{B-0-pos}
B_{t\theta}^{(0)} = -\frac{\frac{\rho_{+}^{2}\rho_{-}^{2}}{R^{3}}}{\sqrt{\left(1+\frac{\rho_{+}^{2}}{R^{2}}\right)\left(1+\frac{\rho_{-}^{2}}{R^{2}}\right)}}
\end{equation}
By transforming back to the original coordinate system or to the canonically normalized one via  \eqref{canon_norm}, we arrive at a conclusion. In \cite{Giveon:2024sgz}, the critical value of the B-field at the origin \eqref{B^0_p} was chosen purely to ensure compatibility between the excited   $w$-winding long string energy spectrum and the $\mathbb{Z}_w$ twisted sector of a single trace $T\bar{T}$ deformed symmetric orbifold dual $CFT_2$ \eqref{TTform}- a.k.a "harmony with single trace $T\bar{T}$ holography". 
We find that this exact same unique value is strictly demanded by the bulk macroscopic thermodynamics governing the long string emission process.

\section{Negative Deformation}
\label{sec:negative}
In this section, we repeat the calculation of the particle and long string geodesics and emission rates,  for the negatively deformed BTZ black hole.

We consider superstring theory on the deformed BTZ black holes background formed in the near $k$ NS5 branes wrapping $S^1\times \mathbb{T}^4 $ with p fundamental \textit{negative} strings (F1)\footnote{See \cite{Dijkgraaf:2016lym}  and references therein for discussions of negative branes in string theory.} winding $S^1$ with asymptotic radius $R$.
The negative strings carry quantized momentum number $n\in \mathbb{Z}$.
The negatively deformed BTZ background, as presented in \cite{Chakraborty:2023zdd}, is given by:
\begin{equation}
\label{defbtz_n}
ds^2=-{N^2\over 1-{\rho^2\over R^2}}d\tau^2+{d\rho^2\over N^2}+{\rho^2\over 1-{\rho^2\over R^2}}(d\theta-N_\theta d\tau)^2~,\qquad \theta\simeq\theta+2\pi~,
\end{equation}
\begin{equation}    
\label{btautheta_n}
B_{\tau\theta}=
{\rho^2\over r_5}\sqrt{\left(1-{\rho_-^2\over R^2}\right)\left(1-{\rho_+^2\over R^2}\right)}{1\over 1-{\rho^2\over R^2}} + B_{\tau\theta}^{(0)}~,
\qquad B_{\tau\theta}^{(0)}\equiv B_{\tau\theta}(\rho=0)~,
\end{equation}
\begin{equation}
\label{dilaton_n}
e^{2\Phi}={kv\over p}\sqrt{\left(1+{\rho_-^2\over R^2}\right)\left(1-{\rho_+^2\over R^2}\right)}{1\over 1-{\rho^2\over R^2}}~,
\qquad v\equiv {\rm Volume}(\mathbb{T}^4)/\left(2\pi\sqrt{\alpha'}\right)^4~
\end{equation}
where $N^2,N_\theta$ and $r_5$ are given in \eqref{withn}.
Note that the background \eqref{defbtz_n}-\eqref{dilaton_n} is obtained from the positively deformed background \eqref{defbtz}-\eqref{dilaton} via the following transformations\footnote{$E_{BH}\to-E_{BH}$ and $E_{ext}\to-E_{ext}$ as well.}:
\begin{equation}
    R^2\to-R^2,\quad M_{BH}\to-M_{BH}.
\end{equation}
These transformations apply directly to the thermodynamic quantities: the mass $M_{BH}$, angular momentum $J_{BH}$ \eqref{M,J}, entropy $S_{BH}$, temperature $T_{BH}$ and angular velocity $\Omega_{BH}$ \eqref{S,T,OM}, consistently satisfying the first law of thermodynamics \eqref{First Law}.
The energy above extremality for this background is given by:
\begin{equation} \label{neg_E}
E_{BH} = M_{BH} - E_{ext} =   -\frac{Rp}{\alpha'} \left(-1 +\frac{1-\frac{\rho_{-}^{2}+\rho_{+}^{2}}{R^{2}}}{\sqrt{\left(1-\frac{\rho_{-}^{2}}{R^{2}}\right)\left(1-\frac{\rho_{+}^{2}}{R^{2}}\right)}} \right)
\end{equation}
As in the previous section, the properties of the black hole solution are compatible with single trace $T\bar{T}$ holography, but now with respect to a negative deformation coupling $\lambda$, given in terms of $R$ as follow:
\begin{equation}
    \lambda = -\frac{\alpha'}{R^2}<0
\end{equation}
Thus, this framework referred as \textit{negative} single trace $T\bar{T}$ holography \cite{Chakraborty:2023zdd,Chakraborty:2020swe,Giveon:2024sgz,Giveon:2023rsk}.
This compatibility becomes explicit when considering theories with fixed entropy and angular momentum along the $\lambda$-deformation line \eqref{fixSJ}. In these scenarios, the energy of the deformed BTZ black hole matches the spectrum of maximally twisted sector $\mathbb{Z}_p$ of a negative single trace $T\bar{T}$ deformed symmetric orbifold $\mathcal{M}^p/S_p$, where each deformed seed $\mathcal{M}$ carries $1/p$ of the total energy:

\begin{equation}
 E_{BH}(\lambda)/p=\frac{1}{\lambda R}\left(-1+\sqrt{1+2\lambda R\frac{E(0)}{p}+\left(\frac{\lambda RP}{p}\right)^2}\right),\qquad \lambda<0.
\end{equation}
Unlike the positively deformed backgrounds, for a fixed momentum $P$, the spectrum of the deformed theory possesses a maximal energy $E_c$ and a corresponding maximal entropy; beyond which, the energy becomes complex:
\begin{equation}
    E_{BH}(\lambda)\le E_c=\frac{Rp}{\alpha'},\qquad S_{BH}(\lambda) \le S_c=2\pi r_5E_c
\end{equation}
Furthermore, these restrictions dictate that the black hole has a maximal size  $\rho_+=R$,  beyond which the $B$-field and $e^{2\Phi}$ become imaginary everywhere\footnote{$\rho_-<R$ for any value of $\rho_+$. In the maximal case, $\rho_-\to 0$ to ensure the angular momentum remains finite \cite{Giveon:2023rsk}.}.

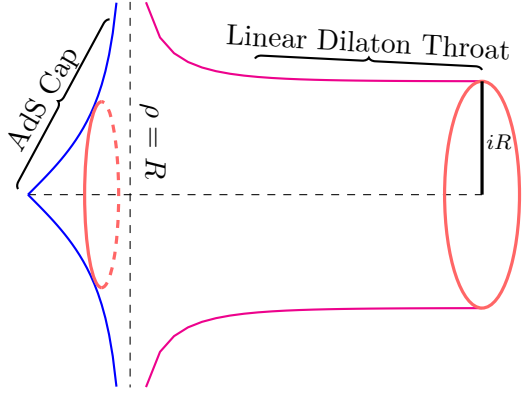
\begin{figure}[h]
    \centering
   \begin{tikzpicture}
[scale = 1.5,domain=0:4]
\draw[thin,dashed,color=black]    plot (\x,0) ;     
  \draw[black,thin,dashed] (0.9,1.7) -- (0.9,-1.7) ;
  % \x r means to convert '\x' from degrees to _r_adians:
  \draw[decorate,decoration={brace}, thick,rotate = 60] (0,0.1)--(1.7,0.15) node[rotate = 60] at (0.8,0.3)  {AdS Cap} ;
  
   \draw[decorate,decoration={brace}, thick] (2,1.2)--(4,1.1) node[rotate = -3] at (3,1.35)  {Linear Dilaton Throat} ;
   \draw[thick,color=blue,domain=0:0.78]   plot (\x,{sqrt((sinh(\x))^2/(1-(sinh(\x))^2))})   ;
  \draw[thick,color=blue,domain=0:0.78]   plot (\x,{-sqrt((sinh(\x))^2/(1-(sinh(\x))^2))}) ;

  \draw[thick,color=magenta,domain=1.04:4]   plot (\x,{sqrt(-(sinh(\x))^2/(1-(sinh(\x))^2))})   ;
  \draw[thick,color=magenta,domain=1.04:4]   plot (\x,{-sqrt(-(sinh(\x))^2/(1-(sinh(\x))^2))}) ;
  
  %\draw[color=red!60 , very thick](1.5,0) ellipse (0.17 and 0.69230769230); %Full ellipse
  
   \draw[color=red!60 , very thick] (0.649208,0.82) arc [
      start angle = 90,
      end angle = 270,
      x radius= 0.15 ,
      y radius = 0.82 
  ];
  \draw[color=red!60 , very thick, dashed] (0.649208,-0.82) arc [
      start angle = 270,
      end angle = 450,
      x radius= 0.15 ,
      y radius = 0.82 
  ];
  
  \draw[color=red!60, very thick](4,0) ellipse (0.33 and 1.00);

  \draw[black, very thick] (4,0) -- (4,1.00);
  \node[right] at (3.95,0.47055) {\scriptsize $iR$};

  \node[rotate = -90] at (1.1,0.47055) {$\rho=R$};
  % undeformed case 
  %\draw[thick,color=orange,domain=0:1.2]   plot (\x,{(\x)});
   %\draw[thick,color=orange,domain=1.2:1.6,dashed]   plot (\x,{\x});
   
  %\draw[thick,color=orange,domain=0:1.2]   plot (\x,{-(\x)}); 
  %\draw[thick,color=orange,domain=1.2:1.6,dashed]   plot (\x,{-(\x)});
  
  %\draw[color=green!60 , very thick](1.2,0) ellipse (0.17 and 1.2);
\end{tikzpicture}  
    \caption{Negative $\lambda$-deformed global $AdS_3$ spatial geometry: the near and far regimes are separated by a gravitational naked singularity at $\rho=R$. The \textit{blue} region is the real sector where $\rho<R$, while the region 
    beyond the singularity at $\rho>R$ becomes purely imaginary, marked in \textit{purple}. This behavior reflects the signature flip in the compact azimuthal direction; the temporal direction also experiences a signature flip  (not depicted). 
    }
  
    \label{fig2}
\end{figure}

A crucial property of the negatively deformed background is the separation between the IR and UV regimes by a \textit{naked singularity} at $\rho = R$, which acts as a UV cutoff.
In the asymptotically flat linear dilaton  regime beyond the singularity $\rho>R$, the metric signature of the temporal $\tau$ and the angular $\theta$ coordinates flips. Consequently, this region contains closed timelike curves (CTCs). These characteristics illustrated in the spatial geometry of the negatively deformed global $AdS_3$ space depicted in Figure \ref{fig2}.

In the following subsections, we will analyze the dynamics of particles and winding string via  their respective worldline and worldsheet actions in this negatively deformed BTZ background. In particular, we will calcualte the emission rates of quanta of particles and $w$-winding long strings using the tunneling and WKB approaches, following the same methodology established in the previous section.

\subsection{Particle Geodesics}
\label{sec:neg_particle}
We now consider the worldline action \eqref{worldline} in the presence of the negatively deformed background.
The geodesics of a particle in this background, governed by the metric in \eqref{defbtz_n}, are determined by the conversation laws for a particle carrying energy $E$ and angular momentum $L$:
\begin{equation} \label{Eptcl_n}
E=-p_{t}=G_{\mu\nu}k_{t}^{\mu}u^{\nu}=\frac{1}{r_{5}\left(1-\frac{\rho^{2}}{R^{2}}\right)}\left(\frac{\rho^{2}-\rho_{+}^{2}-\rho_{-}^{2}}{R^{2}}\dot{t}+\frac{\rho_{-}\rho_{+}}{R}\dot{\theta}\right)
\end{equation}
\begin{equation} \label{Lptcl_n}  L=p_{\theta}=G_{\mu\nu}k_{\theta}^{\mu}u^{\nu}=\frac{1}{r_{5}\left(1-\frac{\rho^{2}}{R^{2}}\right)}\left(\rho^{2}\dot{\theta}-\frac{\rho_{-}\rho_{+}}{R}\dot{t}\right)
\end{equation}
along with the Hamiltonian constraint \eqref{H_constr} and the canonical momentum in \eqref{p_mu}.
Solving the geodesic equations for particle dynamics in the negatively deformed BTZ background \eqref{defbtz_n}, yields the following
radial equation in terms of $E,L$ which relies on the effective square mass $\tilde{m}_\text{eff}^2$:
\begin{equation} \label{eq_rptcl_n}
    \dot{y}^2=4\tilde{m}_{\text{eff}}^2\left(-y^2+\frac{\tilde{\alpha}_d}{\tilde{m}_{\text{eff}}^2} y+\frac{\tilde{\gamma}_d}{\tilde{m}_{\text{eff}}^2}\right)
\end{equation}
where:
\begin{equation} \label{mtilde_eff}
    y=\rho^2,\quad \tilde{m}_{\text{eff}}^{2}=m^{2}+\frac{E^{2}R^{2}-L^{2}}{R^{2}}
\end{equation}
and,
\begin{equation}
    \tilde{\alpha}_{d}(\tilde{m}_{\text{eff}}^2)=E^{2}R^{2}-L^{2}+\tilde{m}_{\text{eff}}^{2}\left(\rho_{+}^{2}+\rho_{-}^{2}\right)+\frac{E^{2}R^{2}\left(\rho_{+}^{2}+\rho_{-}^{2}\right)-2\rho_{-}\rho_{+}ERL}{R^{2}}
\end{equation}
\begin{equation}
    \tilde{\gamma}_{d}(\tilde{m}_{\text{eff}}^2)=\left(\rho_{+}^{2}+\rho_{-}^{2}\right)L^{2}-2ERL\rho_{-}\rho_{+}-\tilde{m}_{\text{eff}}^{2}\rho_{+}^{2}\rho_{-}^{2}+\frac{E^{2}R^{2}-L^{2}}{R^{2}}\rho_{+}^{2}\rho_{-}^{2}
\end{equation}
Here, we used tilde notation to distinguish the parameters of the solution under a deformation from those of the positively deformed solution given in the equations \eqref{meff}-\eqref{B,C}.
The structure of the radial equation \eqref{eq_rptcl_n} is identical to \eqref{eq_rptcl} and thus its solution maps directly to \eqref{rho_sol} under the following parameter replacements:
\begin{equation} \label{transtonega}
    m^2_\text{eff}\to \tilde{m}^2_\text{eff},\qquad\alpha_d\to \tilde{\alpha}_d,\qquad \gamma_d\to \tilde{\gamma}_d,\qquad \beta_d\to \tilde{\beta}_d
\end{equation}
where:
\begin{equation}
    \tilde{\beta}_{d}\left(\tilde{m}_{\text{eff}}^{2}\right)=\tilde{\alpha}_{d}^{2}\left(\tilde{m}_{\text{eff}}^{2}\right)+4\tilde{m}_{\text{eff}}^{2}\\\tilde{\gamma}_d\left(\tilde{m}_{\text{eff}}^{2}\right). 
\end{equation}
Furthermore, using \eqref{Eptcl_n} and \eqref{Lptcl_n} the temporal and angular coordinates can be expressed explicitly in terms of $\rho^2$:
\begin{multline*}
t=t_{0}\pm\frac{r_{5}}{2\left(\rho_{+}^{2}-\rho_{-}^{2}\right)}\left[\rho_{+}\sqrt{1-\frac{\rho_{+}^{2}}{R^{2}}}\tilde{f}^{-}\left(\rho^{2}-\rho_{+}^{2},\tilde{B}_{+},\tilde{C}_{+}\right)+\rho_{-}\sqrt{1-\frac{\rho_{-}^{2}}{R^{2}}}\tilde{f}^{+}\left(\rho^{2}-\rho_{-}^{2},\tilde{B}_{-},\tilde{C}_{-}\right)\right] \\
\mp\frac{E}{R}\tilde{h}\left(\rho^{2}-\rho_{+}^{2},\tilde{B}_{+},\tilde{C}_{+}\right)
\end{multline*} 
\begin{multline}   \label{theta_sol_n}
\theta=\theta_{0}\pm\frac{r_{5}}{2\left(\rho_{+}^{2}-\rho_{-}^{2}\right)}\left[\rho_{-}\sqrt{1-\frac{\rho_{+}^{2}}{R^{2}}}\tilde{f}^{-}\left(\rho^{2}-\rho_{+}^{2},\tilde{B}_{+},\tilde{C}_{+}\right)-\rho_{+}\sqrt{1-\frac{\rho_{-}^{2}}{R^{2}}}\tilde{f}^{-}\left(\rho^{2}-\rho_{-}^{2},\tilde{B}_{-},\tilde{C}_{-}\right)\right] \\ 
\mp\frac{L}{R^{2}}\tilde{h}\left(\rho^{2}-\rho_{+}^{2},\tilde{B}_{+},\tilde{C}_{+}\right)
\end{multline}
with,
\begin{align} 
\tilde{f}^{\pm}\left(x,\tilde{B},\tilde{C}\right) &= \log\left(\frac{\frac{x}{R^{2}}}{2\tilde{C}\sqrt{-m_{\text{eff}}^{2}x^{2}+\tilde{B}x+\tilde{C}^{2}}\pm\tilde{B}x\pm2\tilde{C}_{+}^{2}}\right)\\ \tilde{h}\left(x,\tilde{B},\tilde{C}\right) &= \frac{r_{5}}{2\sqrt{m_{\text{eff}}^{2}}}\arctan\left(\frac{-2m_{\text{eff}}^{2}x+\tilde{B}}{2\sqrt{m_{\text{eff}}^{2}}\sqrt{-m_{\text{eff}}^{2}x^{2}+\tilde{B}_{+}x+\tilde{C}^{2}}}\right)
\end{align}
and,
\begin{equation} \label{B,C}
    \tilde{B}_{\pm}=E^{2}R^{2}-L^{2}\mp m_{\text{eff}}^{2}\left(\rho_{+}^{2}-\rho_{-}^{2}\right)-\frac{E^{2}R^{2}\left(\rho_{+}^{2}+\rho_{-}^{2}\right)-2ERL\rho_{-}\rho_{+}}{R^{2}},\quad\tilde{C}_{\pm}=\left(ER\rho_{\pm}-L\rho_{\mp}\right)\sqrt{1-\frac{\rho_{\pm}^{2}}{R^{2}}}.
\end{equation}
The qualitative behavior of probe particles in the negatively $\lambda$ -deformed BTZ  geometry is essentially  inverted with respect to the positive case. Timelike particles $m^2<\tilde{m}_c^2$  will reach the asymptotic flat regime  when $L>ER$ (as oppose to $ER>L$ in the positive case). This is reflected through the critical mass squared, which is now given by:
\begin{equation}
\tilde{m}_c^2 = -m_c^2 = \frac{L^2-E^2R^2}{R^2}
\end{equation}
In other cases, heavier particles reach a finite maximal radius given by \eqref{rhomax}\footnote{Evaluated using the mapped parameters \eqref{transtonega}}. They then either fall back and terminate at the singularity or, if they possess sufficient angular momentum, oscillate in a region bounded below by a non-zero minimal radius, exactly as discussed previously for the positively deformed case.

Interestingly, the existence of the UV wall at $\rho = R$ does not strictly forbids to heavy and light particles from entering the radial regime beyond $\rho>R$, due to the analytical smoothness of the radial solution \eqref{rho_sol}. However, the $t,\theta$ coordinate velocities completely freeze out at $\rho=R$. Therefore, unlike the positively deformed case, the presence of the UV wall effectively prevents particles emitted from the horizon, from ever reaching asymptotic boundary at $\rho,t\to\infty$.

 \subsection{Particle Emission from the Deformed BTZ}
As in the positively deformed BTZ case, the negatively deformed BTZ background also suffer from a coordinate singularity at the outer horizon $\rho=\rho_+$ when using the original coordinate system \eqref{defbtz_n}.
Therefore, we will again transform to the regular Gullstrand–Painlevé coordinate system $(\frak{t},\rho,\varphi)$ in order to properly treat the tunneling dynamics near the horizon. For negatively deformed BTZ geometry, the coordinate transformations take the form:
\begin{equation} \label{PG_n}
    dt=d\mathfrak{t}-\frac{R}{r_{5}}\sqrt{\frac{1}{N^{2}}\left(\frac{1}{N^{2}}-\frac{1}{f\left(\rho\right)}\right)\left(1-\frac{\rho^{2}}{R^{2}}\right)}d\rho,\quad d\theta=d\varphi-N_{\theta}\sqrt{\frac{1}{N^{2}}\left(\frac{1}{N^{2}}-\frac{1}{f\left(\rho\right)}\right)\left(1-\frac{\rho^{2}}{R^{2}}\right)}d\rho
\end{equation}
for an arbitrary smooth function $f(\rho)$, giving rise to the following metric:

\begin{equation}
ds^{2}=\frac{d\rho^{2}}{f\left(\rho\right)}+\frac{1}{1-\frac{\rho^{2}}{R^{2}}}\left(-\frac{r_{5}^{2}}{R^{2}}N^{2}d\mathfrak{t}^{2}+\frac{2r_{5}}{R}\sqrt{N^{2}\left(\frac{1}{N^{2}}-\frac{1}{f\left(\rho\right)}\right)\left(1-\frac{\rho^{2}}{R^{2}}\right)}d\rho d\mathfrak{t}+\rho^{2}\left(d\varphi-\frac{r_{5}}{R}N_{\theta}d\mathfrak{t}\right)^{2}\right).
\end{equation}
To compute the probability amplitude for the emission of a   particle carrying energy $E$ and angular momentum $L$ from the negatively deformed BTZ, we again integrate along the radial direction from the initial horizon $\rho_{in}=\rho_+(M_{BH},J_{BH})$ to to the final remnant horizon $\rho_{out}=\rho_+(M_{BH}-E,J_{BH}-L)$ using the Hamiltonian worldline path integral \eqref{I_ptcl}. In the analysis below, we assume that the outer horizon before the emission satisfies $\rho_+<R$:
 \begin{equation} \label{ImI_n}
 \Im\mathcal{I}_{\text{ptcl}}=\Im\left(\int_{\rho_{in}}^{\rho_{out}}d\rho\,p_{\rho}\right)=\Im\left(\int_{\rho_{in}}^{\rho_{out}}d\rho\,\int_{0}^{p_{\rho}}dp_{\rho}'\right)  
 \end{equation}
By utilizing the conserved quantities \eqref{Eptcl_n} and \eqref{Lptcl_n} evaluated in the Gullstrand–Painlevé coordinates alongside the Hamiltonian constraint \eqref{H_constr}, we obtain the radial momentum, which in the near horizon limit is given by:
\begin{equation}
    p_{\rho}\sim \frac{r_{5}\sqrt{1-\frac{\rho_{+}^{2}}{R^{2}}}\left(ER\rho_{+}-L\rho_{-}\right)}{\left(\rho-\rho_{+}\right)\left(\rho_{+}^{2}-\rho_{-}^{2}\right)}
\end{equation}
We note that for maximal black holes with $\rho_+=R$ and $\rho_-=0$, the radial momentum $p_\rho$ completely vanishes, thus constitutes a breakdown point for the semiclassical tunneling approach in the maximally negatively deformed background.

Taking the differential of the radial momentum yields:
\begin{equation} \label{dpr_n}
    dp_{\rho} = -\frac{Rr_{5}\rho_{+}\sqrt{1-\frac{\rho_{+}^{2}}{R^{2}}}}{\left(\rho-\rho_{+}\right)\left(\rho_{+}^{2}-\rho_{-}^{2}\right)}\left(d\tilde{M}_{BH}-\Omega_{BH}d\tilde{J}_{BH}\right) 
\end{equation}
where $d\tilde{M}_{BH}$ and $d\tilde{J}_{BH}$
are the differential shifts in the black hole's mass and angular momentum during the particles emission process.
Substituting \eqref{dpr_n} into the imaginary part of the Hamiltonian path integral \eqref{ImI_n}, performing the analytical continuation into the complex plane, and applying the Sokhotski–Plemelj theorem directly yields the exact same universal thermodynamic relation previously established in \eqref{ImI}. In doing so, we have used the fact that in the negatively deformed case, the black holes temperature  takes the form\footnote{The maximal black hole at $\rho_+=R$ has an infinite temperature; in this limit the emergent geometry approaches the $SL(2,\mathbb{R})_k/U(1)$ geometry at $\rho>R$ and a 2D cosmology for $\rho<R$ \cite{Giveon:2023rsk}. }:
\begin{equation}
    T_{BH}=\frac{\rho_{+}^{2}-\rho_{-}^{2}}{2\pi r_{5}R\rho_{+}\sqrt{1-\frac{\rho_{+}^{2}}{R^{2}}}}.
\end{equation}
This implies, in particular, that the exponential emission probability maintains the universal form \eqref{Gamma2} for the deformed BTZ with negative coupling.
Furthermore, a similar WKB approach can also  be applied to the negatively deformed BTZ case. Expanding the entropy change \eqref{DeltaS}- expressed in the Cardy form \eqref{Entropy} 
to linear order with respect to the the emitted particle's energy and angular momentum, demonstrates complete consistency with the black hole first law of thermodynamics.

\subsection{Winding Long String Emission from the Deformed BTZ}
\label{sec:neg_strings}

Just as in the positively deformed case, the geodesics of long strings in the negatively deformed BTZ are derived directly from the particle trajectories: \eqref{rho_sol},\eqref{theta_sol_n}, and are extended to the case of strings by applying the spectral flow operation \eqref{spectralflow}.
The fact that the coordinates $t,\theta$ flip their signatures beyond the UV wall at $\rho>R$ does not affect the spectral flow transformations. This resilience relies on the fact that closed long strings winding the compact coordinate, together with the time coordinate possess winding symmetry in the far UV limit as well as in the near IR limit.

As opposed to particles, the spectral flow operation allows a long string in the negatively deformed BTZ geometry, to reach the asymptotic boundary at $t,\rho \to \infty$. A long string, initiated at the horizon $\rho=\rho_+$, moving forward in time and reach the UV wall at $\rho=R$ with a rate $w>0$, allowing it to pass into the $\rho>R$ regime in finite  time. However, for the coordinate time to keep grow forward along with the radial direction, the string must maintain minimal a winding rate above a specific threshold:

\begin{equation} \label{wthreshold}
\lim_{\rho\to\infty}|\dot{t}|=r_5 E<|w|.
\end{equation}

The excitation energy of long strings in the negatively deformed BTZ will be treated as follow: we denote $E_{w,string}$ as the conjugate energy excitation momenta of the temporal canonical normalized coordinate $t$ \eqref{canon_norm} (which has a positive signature asymptotically), and $L_{string}$ as the angular momentum excitation number conjugate to the angular coordinate $\theta$ (which has a negative signature asymptotically), as was done explicitly in \cite{Giveon:2024sgz}.
As discussed previously, by using only the asymptotic data, we can fix a non-trivial value for the B-field at the origin to ensure harmony with negative single trace $T\bar{T}$ spectrum, that is:
\begin{equation} \label{B^0_n}
     B_{tx}^{(0)} = \frac{\frac{\rho_-^2\rho_+^2}{R^4}}{\sqrt{\left(1-\frac{\rho_-^2}{R^2}\right)\left(1-\frac{\rho_+^2}{R^2}\right)}} \Rightarrow B_{\tau \theta}^{(0)} =\frac{R^2}{r_5}\frac{\frac{\rho_-^2\rho_+^2}{R^4}}{\sqrt{\left(1-\frac{\rho_-^2}{R^2}\right)\left(1-
     \frac{\rho_+^2}{R^2}\right)}} 
\end{equation}
This choice, provides that $w$-excited long string spectrum coincides with that of a $\mathbb{Z}_w$ twisted sector of a \textit{negative} single trace $T\bar{T}$ deformed symmetric orbifold $\mathcal{M}^p/S_p$:

\begin{equation}
 E_{w,total}(\lambda)=\frac{w}{\lambda R}\left(-1+\sqrt{1+\frac{2\lambda \mathcal{E}_{w,total}}{w}+\left(\frac{\lambda L_{string}}{w}\right)^2}\right),\quad \lambda = -\frac{\alpha'}{R^2}<0  
\end{equation}
where $E_{w,total}$ and $\mathcal{E}_{w,total}$ are defined and described around \eqref{total deformed}-\eqref{undef tot}.
Just as with black hole itself, the energy of winding long strings is bounded by a maximal excitation energy, beyond which it becomes complex:
\begin{equation}
 E_{w,string} \le \frac{Rw}{\alpha'}.
\end{equation}

In the presence of long string states, the thermodynamic environment of the black hole is described by the grand canonical ensemble, governed by the first law of thermodynamics \eqref{grand_canon_1}
with a chemical potential given by:
\begin{equation} \label{chemical_neg}
    \mu_{BH} = -\frac{1}{\alpha'}\left(\frac{\rho_{+}^{2}}{R}\sqrt{\frac{1-\frac{\rho_{-}^{2}}{R^{2}}}{1-\frac{\rho_{+}^{2}}{R^{2}}}}+B_{t\theta}^{\left(0\right)}\right)
\end{equation}
The emission probability for these long strings is computed in a free-falling observer frame, utilizing the Gullstrand–Painlevé coordinate system \eqref{PG_n}. These coordinates introduce non-trivial B-field components $B_{\frak{t}\rho},B_{\varphi\rho}$, which can be gauged away by a proper gauge fixing e.g. \eqref{gaugefix}, while maintaining $B_{t\theta}=B_{\mathfrak{t}\varphi}$.
The imaginary part of the long strings' Hamiltonian path integral action \eqref{Istr} eventually reduces to the universal emission rate \eqref{Gamma2}, receiving contributions exclusively from the radial momentum $p_{\rho}$. To recapitulate the calculation, we consider the radial momentum obtained by solving Hamiltonian constraint equation \eqref{H_st_con}-\eqref{momentums}. To leading order in the near horizon limit, it is given by:
\begin{equation}
    p_{\rho}\sim\frac{r_{5}\sqrt{1-\frac{\rho_{+}^{2}}{R^{2}}}}{\left(\rho-\rho_{+}\right)\left(\rho_{+}^{2}-\rho_{-}^{2}\right)}\left(E_{w,string}R\rho_{+}-L_{string}\rho_{-}+\frac{w R\rho_{+}}{\alpha'}\left(\frac{\rho_{+}^{2}}{R}\sqrt{\frac{1-\frac{\rho_{-}^{2}}{R^{2}}}{1-\frac{\rho_{+}^{2}}{R^{2}}}}+B_{t\theta}^{\left(0\right)}\right)\right).
\end{equation}
As with particle emission, the maximal black hole constitutes a breakdown point for the semiclassical tunneling process; however, for long strings, the radial momentum reaches a nonzero value rather than vanishing entirely. 

Substituting the radial momentum into the Hamiltonian action and preforming the integration across the classically forbidden regime- between black hole horizon before and after the emission process, in both tunneling and WKB approaches, ultimately yields the same universal result \eqref{ImIst} found in the positively deformed case above, as well as in the undeformed case in \cite{Martinec:2023plo}.
The probability for long string emission is highly suppressed for large winding $w$ states. Consequently, most of the emitted strings will never reach the asymptotic boundary; only a tiny fraction of the emitted states will carry winding large enough to overcome the winding rate threshold \eqref{wthreshold} and make it to the asymptotic boundary at $t\to \infty$.

Repeating the analysis from \eqref{DeltaS}-\eqref{dS_BH_sol}, which expands the change in the entropy after long string quanta emission within the WKB paradigm, reveals 
agreement with the first law of thermodynamics in the grand canonical ensemble \eqref{grand_canon_1}, provided a specific choice for the chemical potential:

\begin{equation} \label{pref_chem_neg}
    \mu_{BH}=-\frac{1}{\alpha'}\frac{\rho_{+}^{2}}{R\sqrt{\left(1-\frac{\rho_{-}^{2}}{R^{2}}\right)\left(1-\frac{\rho_{+}^{2}}{R^{2}}\right)}}
\end{equation}
Comparing this required potential \eqref{chemical_neg}, yields a unique, thermodynamically favored value for the B-field at the origin:
\begin{equation} \label{B-0-neg}
B_{t\theta}^{\left(0\right)}=\frac{\frac{\rho_{+}^{2}\rho_{-}^{2}}{R^{3}}}{\sqrt{\left(1-\frac{\rho_{-}^{2}}{R^{2}}\right)\left(1-\frac{\rho_{+}^{2}}{R^{2}}\right)}}
\end{equation}
By transforming back to the original or canonically normalized coordinate system \eqref{canon_norm}, we see that this thermodynamic value for the B-field at the origin coincides with the value in \eqref{B^0_n}. As shown in \cite{Giveon:2024sgz}, this specific value is what guarantees the compatibility of the excited $w$-winding long string energy spectrum to that of the $\mathbb{Z}_w$ twisted sector (and untwisted sector for $w=1$) of a single trace $T\bar{T}$ deformed symmetric orbifold.

\section{Extension To General Deformed BTZ$\times S^1$}
\label{sec:exten_TJJT}
The analysis above for particle and winding long string emission from deformed BTZ backgrounds can be generalized directly to a wider class of linear dilaton, asymptotically flat deformed BTZ black hole solutions.
Such solutions are obtained by the action of truly-marginal deformations on the $SL(2,\mathbb{R})_k\times U(1)$
WZW worldsheet theory \cite{Chakraborty:2019mdf}.
In the dual theory, they amount to linear combination of irrelevant deformations including $T\bar{T},J\bar{T}$ and $T\bar{J}$ operators, where $J(y)$ and $\bar{J}(y)$ are extra left and right-moving conserved currents.
Consequently, the deformed theory is a symmetric product $\mathcal{M}^p/S_p$  
where each seed CFT$_2$ $\mathcal{M}$ carrying the central charge $c=6k$, is deformed by sequential operations of the $T\bar{T},J\bar{T},T\bar{J}$ operators, thus referred as a "single trace $T\bar{T}+J\bar{T}+T\bar{J}$ deformation".

From the spacetime effective supergravity perspective, the deformed background is obtained by the decoupling limit of $k$ NS5 branes wrapping $S^1_x\times S^1_y\times \mathbb{T}_3$ and $p$-fundamental strings (F1) winding $S^1_x$.
This gives rise to a $(\lambda,\epsilon_\pm)$-deformed $AdS_3\times S_y^1$ background, extending the solution space from a line of $\lambda$-deformed solutions to a larger space of solutions spanned by the parameters  $\lambda,\epsilon_{\pm}$. The dimensionless couplings $\epsilon_\pm$ are associated with the current-current deformations on the $U(1)$ piece of the worldsheet theory and are holographically mapped to the dimensionful coupling of the $J\bar{T}$ and $T\bar{J}$ operators acting on the boundary CFT$_2$.

In \cite{Chakraborty:2019mdf}, the solution for $(\lambda,\epsilon_\pm)$-deformed massless BTZ was found, and the energy spectrum of excited $w$-winding long strings above the background energy threshold was derived. This exactly matches the spectrum for the $\mathbb{Z}_w$ twisted sector of a single trace $T\bar{T}+J\bar{T}+T\bar{J}$ deformed symmetric orbifold, given explicitly by:
\begin{equation} \label{spectJT}
    E_{w,total}R = \frac{1}{2a}\left(-b-\sqrt{b^2-4ac}\right)
\end{equation}
with $a,b,c$ given by\footnote{We use the conventions of eq.(22) in \cite{Giveon:2025cyk}}:
\begin{equation} \label{E_wtot TT TJ JT}
\begin{aligned}
a &= -\frac{\Psi}{2},\qquad \Psi=\lambda-(\epsilon_-+\epsilon_+)^2 \\
b &= -w+\left(\epsilon_+^2-\epsilon_-^2\right)n_x-\sqrt{\alpha'}\left(q_R\epsilon_--q_L\epsilon_+\right) \\
c &= \frac{1}{2}\left(\lambda + 
(\epsilon_--\epsilon_+)^2\right)n_x^2+\sqrt{\alpha'}\left(q_R\epsilon_--q_L\epsilon_+\right)n_x+w\mathcal{E}_{w,total}.
\end{aligned}
\end{equation}
Here, $w$ is the winding number of the long string around the $S_x^1$ circle associated with the azimuthal direction of the BTZ black hole, $n_x$ is the excited angular momentum number around $S_x^1$. The parameters
$q_L,q_R$ are the left and right charges along the extra circle $S_y^1$, which are given in terms of the excited winding charge $w_y$ and excited angular momentum charge $n_y$ according to:
\begin{equation}
   q_{L,R} = \frac{w_yR_y}{\alpha'}\pm \frac{n_y}{R_y}
\end{equation}
where $R_y$ is an arbitrary radius of the $S_y^1$ circle, such that $y\sim y+2\pi R_y$. The parameter $R_x$ is the asymptotic radius of $S^1_x$ circle which expressed via the deformation parameters as:
\begin{equation} \label{R_X}
    R=\sqrt{\frac{\alpha'}{\lambda-4\epsilon_-\epsilon_+}}
\end{equation}
and $\mathcal{E}_{w,total}
$ is the total energy of the undeformed theory \eqref{undef tot}.
The $\epsilon_\pm\to 0$ limit reduces to the long string spectra above $\lambda$-deformed BTZ considered in the previous sections \eqref{lambda def},\eqref{TTform}-\eqref{lim_undef}, matching the spectrum of a $\mathbb{Z}_w$ twisted sector of a single trace $T\bar{T}$ deformed symmetric orbifold.
The spectrum \eqref{E_wtot TT TJ JT} is completely free from complex energies one if $\Psi\ge 0$. In the case where $\Psi<0$, the system has a maximal energy value beyond which the energy becomes complex; this extends the behavior we observed for the negative deformation case in Section \ref{sec:negative}:
\begin{equation}
    E_{w,total}\le \frac{w}{\Psi R}.
\end{equation}
Another subspace with pathological spacetime geometry appears when $\lambda-4\epsilon_-\epsilon_+<0$, which gives rise to a complex asymptotic radius $R$ \eqref{R_X} and must be treated carefully in a similar manners as the case of long negative strings above.

In \cite{Giveon:2025cyk}, the asymptotic data of the massless deformed BTZ was used to extend the compatibility of long string excitation spectrum with that of a single trace $T\bar{T}+J\bar{T}+T\bar{J}$ symmetric product, to a massive BTZ backgrounds with vanishing angular momentum and left-right charges.
It was found that, just as in the $\lambda$-deformed BTZ case, there is a uniquely favored value for the asymptotic B-field components. Relying on this, we might fix uniquely the value of the B-field components at the origin once a suitable solution is constructed. Such a supergravity black hole solution requires an energy above extremality that can be cast into to the form of \eqref{E_wtot TT TJ JT} for theories with fixed entropy and charges along the $(\lambda,\epsilon_\pm)$ parameter space\footnote{In such a theory, the energy is given by \eqref{E_BH+} with the following identification $E_{w,total}\to E_{BH},\quad\mathcal{E}_{w,total}\to {r_5 M_{BTZ}},\quad n\to J_{BTZ},\quad q_{L,R}\to Q_{L,R},\quad w\to p$. For purely massive background, $J_{BTZ},Q_{L,R}\to0$. } \eqref{fixedcharges}. 

It was argued in \cite{Apolo:2021wcn} that a sequence of solution-generating transformation on the undeformed background, tri-TsT transformations, gives rise to a large space of deformed BTZ$\times S^1$ solutions having  total winding charge $p$, carrying mass $M_{BH}$, angular momentum $J_{BH}$ and charges $Q_{L,R}$. The spectrum of these solutions consolidate with the maximal twisted $\mathbb{Z}_p$ spectrum of single trace $T\bar{T}+J\bar{T}+T\bar{J}$ deformed theories. However, establishing a clear matching between the parameters of \cite{Apolo:2021wcn}
and the conventions utilized throughout this note, are left for future work.

Below, we consider a general deformed BTZ setup without specifying the exact solution and its parameters, beyond the thermodynamical quantities and the deformation parameters.
In the context of winding long string emission, the gradual evaporation process is governed by the grand canonical ensemble. For the doubly charged black hole solutions of interest, the first law of thermodynamics is modified from \eqref{grand_canon_1} to include the effects coming from the extra left and right charges $Q_{L,R}$:
\begin{equation} \label{Grand-Entropy-JT}
     dS_{BH} = \frac{1}{T_{BH}}\left(dE_{BH}-\Omega_{BH}dJ_{BH}-\mu_{BH}dp-\mu_L dQ_L - \mu_R dQ_R\right)
\end{equation}
where $\mu_{L,R}$ are the corresponding left and right chemical potentials.

For an emitted long string carrying a winding $w$, excitation energy $E_{w,string}$, angular momentum $L_{string}$ and left/right charges $\mathcal{Q}_{L,R}$, one expects that the probability of tunneling through the horizon barrier to still obey the universal law\footnote{We do not provide a proof that relies on the direct derivation of the spectral flow for these theories; instead, we assume in the upcoming  analysis that the universal result \eqref{Gamma2}  holds here as well. } \eqref{Gamma2}. That is, the emission probability governed predominantly by the change in the black hole's entropy, which for the doubly charged deformed BTZ is expressed by:
\begin{equation} \label{DeltaSJT}
    \Delta S_{BH} = S_{BH}(E_{BH}-E_{w,total},J_{BH}-L_{string},p-w,Q_{L,R}-\mathcal{Q}_{L,R})-S_{BH}(E_{BH},J_{BH},p,Q_{L,R})
\end{equation}
Using the $AdS/CFT$ dictionary, one can write the entropy of such black holes in a Cardy-like form:

\begin{equation} \label{CardyJT}
    S_{BH}=2\pi\sqrt{\frac{c}{6}}\left(\sqrt{E_{L}\left(\lambda,\epsilon_{\pm}\right)+F\left(\lambda,\epsilon_{\pm}\right)}+\sqrt{E_{R}\left(\lambda,\epsilon_{\pm}\right)+F\left(\lambda,\epsilon_{\pm}\right)}\right)
\end{equation}
where $c=6kp$ and $F$ is a function of the deformation parameters $(\lambda,\epsilon_\pm)$, the left/right charges $Q_{L,R}$ and left/right deformed energies $E_{L,R} (\lambda,\epsilon_\pm)$ \eqref{ELR}, given by:
\begin{multline}
    F\left(\lambda,\epsilon_{\pm}\right)=\frac{\lambda}{p}E_{L}\left(\lambda,\epsilon_{\pm}\right)E_{R}\left(\lambda,\epsilon_{\pm}\right)+ \\ \frac{1}{p}\left[\sqrt{\alpha'}\left(\epsilon_{-}Q_{R}E_{R}\left(\lambda,\epsilon_{\pm}\right)+\epsilon_{+}Q_{L}E_{L}\left(\lambda,\epsilon_{\pm}\right)\right)-\left(\epsilon_{-}E_{R}\left(\lambda,\epsilon_{\pm}\right)+\epsilon_{+}E_{L}\left(\lambda,\epsilon_{\pm}\right)\right)^{2}\right]
\end{multline}
Looking at theories with fixed entropy along the $(\lambda,\epsilon_\pm)$ parameter space, similar to what was done in \eqref{fixSJ} for the $\lambda$-deformed BTZ, one obtains the flow equation which defines the deformed spectrum above \eqref{spectJT}-\eqref{E_wtot TT TJ JT}:
\begin{equation} \label{Flow+TJJT}  E_{L,R}\left(\lambda,\epsilon_{\pm}\right)+F\left(\lambda,\epsilon_{\pm}\right)=E_{L,R}\left(0\right)
\end{equation}
subject fixed angular momentum and charges\footnote{Analyzing $J\bar{T},T\bar{J}$ deformed CFTs usually give rise to additional flow equations for the left and right charges see e.g. eq.(6.18) in \cite{Chakraborty:2018vja}. However, here we retain the same definition of left and right  energies \eqref{ELR} as in the uncharged $\lambda$-deformed BTZ. Presenting the flow equation in the form of \eqref{Flow+TJJT}, allows us to absorb the flow equation of $Q_{L,R}$ and treat it as a fixed quantity. In this scenario, $Q_{L,R}(0)$ simply represent the deformed black hole ADM charges. Similar ideas were implemented for the black hole solutions presented in \cite{Apolo:2021wcn}.} 
\begin{equation} \label{fixedcharges}
    Q_{L,R}\left(\lambda,\epsilon_{\pm}\right)=Q_{L}\left(0\right),\qquad J_{BH}\left(\lambda,\epsilon_{\pm}\right)=J_{BH}(0)=J_{BTZ}.
\end{equation}
Substituting the Cardy-like entropy of the deformed theory \eqref{CardyJT} into the change of the entropy upon emitting a charged $w$-winding long string \eqref{DeltaSJT}, evaluating it to linear order in $\frac{E_{w,string}}{E_{BH}},\frac{L_{string}}{J_{BH}},\frac{\mathcal{Q}_{L,R}}{Q_{L,R}},\frac{w}{p}\ll1$,
and following the WKB prescript established in \eqref{dS_BH_sol}, one can determine the insensitive thermodynamic variables by comparing the results directly to the first law \eqref{Grand-Entropy-JT}:
\begin{multline} \label{TBH}
  T_{BH}=\frac{2}{\pi R\sqrt{kp}}\left(\frac{\sqrt{\left(E_{R}+F\right)\left(E_{L}+F\right)}}{\sqrt{E_{L}+F}+\sqrt{E_{R}+F}}\right)\times\\ \left(1+\frac{\lambda}{p}E_{BH}R-\frac{2\left(\epsilon_{+}+\epsilon_{-}\right)}{p}\left(\epsilon_{-}E_{R}+\epsilon_{+}E_{L}\right)+\frac{\sqrt{\alpha'}}{p}\left(\epsilon_{+}Q_{L}+\epsilon_{-}Q_{R}\right)\right)^{-1}  
\end{multline}
\begin{multline} \label{omegaBH}
    \Omega_{BH}=\frac{\pi\sqrt{kp}T_{BH}}{2}\Bigg{(}\frac{1+\frac{\lambda}{p}J_{BH}-\frac{2\left(\epsilon_{-}-\epsilon_{+}\right)}{p}\left(\epsilon_{-}E_{R}+\epsilon_{+}E_{L}\right)-\frac{\sqrt{\alpha'}}{p}\left(\epsilon_{+}Q_{L}-\epsilon_{-}Q_{R}\right)}{\sqrt{E_{R}+F}}- \\ \frac{1-\frac{\lambda}{p}J_{BH}+\frac{2\left(\epsilon_{-}-\epsilon_{+}\right)}{p}\left(\epsilon_{-}E_{R}+\epsilon_{+}E_{L}\right)+\frac{\sqrt{\alpha'}}{p}\left(\epsilon_{+}Q_{L}-\epsilon_{-}Q_{R}\right)}{\sqrt{E_{L}+F}}\Bigg{)}
\end{multline}
\begin{multline}
    \mu_{BH}=-\frac{\pi T_{BH}}{2}\sqrt{\frac{k}{p}}\Bigg{[}\left(E_{BH}R\left[1+\frac{\lambda}{p}E_{BH}R-\frac{2\left(\epsilon_{+}+\epsilon_{-}\right)}{p}\left(\epsilon_{-}E_{R}+\epsilon_{+}E_{L}\right)+\frac{\sqrt{\alpha'}}{p}\left(\epsilon_{+}Q_{L}+\epsilon_{-}Q_{R}\right)\right]-2F\right) \times \\ \left(\frac{\sqrt{E_{L}+F}+\sqrt{E_{R}+F}}{\sqrt{\left(E_{R}+F\right)\left(E_{L}+F\right)}}\right) - 2\left(\sqrt{E_{L}+F}+\sqrt{E_{R}+F}\right)\Bigg{]}
\end{multline}
 \begin{equation} \label{muLR}
     \mu_{L,R}=-\frac{\pi r_{5}T_{BH}}{\sqrt{p}}\left(\frac{\sqrt{E_{L}+F}+\sqrt{E_{R}+F}}{\sqrt{\left(E_{R}+F\right)\left(E_{L}+F\right)}}\right)\epsilon_{\pm}E_{L,R}
 \end{equation}   
Here we choose to present the deformed left and right energies $E_{L,R}$ and $F$ without explicitly noting their dependence on $(\lambda.\epsilon_\pm)$ for simplicity.
Once we find a suitable effective black-string solution with characteristic parameters that allows us to express the ADM extensive quantities $E_{BH},J_{BH},Q_{L,R}$,
we can directly extract the intensive quantities $T_{BH},\Omega_{BH}$ and the thermodynamically favored chemical potentials for the emission process $\mu_{BH},\mu_{L,R}$ using the relations above\footnote{By taking $\epsilon_\pm\to 0$, one smoothly restores the intensive thermodynamic variables of the $\lambda$-deformed BTZ black hole \eqref{S,T,OM}, which has $\rho_{\pm}$ as its characteristic parameters. In particular, one precisely recovers the preferable chemical potential given in \eqref{pref_chem_pos} and \eqref{pref_chem_neg}.}.
Using this data, one can express the angular potential around $S_y^1$ circle, $\Omega_y$ (conjugate to the momentum number $n_y$) and the winding chemical potential $\mu_y$ (conjugate to the winding number $w_y$) via:
\begin{equation} \label{omegamu-y}
    \Omega_y=\frac{\mu_L-\mu_R}{
    R_y},\qquad \mu_y = \frac{R_y(\mu_L+\mu_R)}{\alpha'}.
\end{equation}
These potentials, along with $\Omega_{BH},\mu_{BH}$ are directly related to the metric and B-field components of the 3+1-dimensional solution, evaluated at the horizon \cite{Compere:2007vx}:

\begin{align}
    \Omega_y &=\frac{1}{R_y}\frac{G_{ty}}{G_{yy}}\Big|_{\rho=\rho_+},\qquad \Omega_{BH} =\frac{G_{t\theta}}{G_{\theta \theta}}\Big|_{\rho=\rho_+}  \\ \label{muy}
    \mu_y&=-\frac{R_y}{\alpha'}\zeta^\alpha_{H}B_{\alpha y}|_{\rho=\rho_+}=-\frac{R_y}{\alpha'}\left(B_{t y}-\Omega_{BH}B_{\theta y}\right)|_{\rho=\rho_+}
    \\ \label{muBH}
    \mu_{BH}&=-\frac{1}{\alpha'}\zeta^\alpha_{H}B_{\alpha \theta}|_{\rho=\rho_+}=-\frac{1}{\alpha'}\left(B_{t \theta}+\Omega_yR_yB_{\theta y}\right) |_{\rho=\rho_+}
\end{align}
where the doubly charged 3+1 black hole solution now has the following Killing horizon vector:
\begin{equation}
    \zeta_H^a=\partial_t-\Omega_{BH}\partial_\theta-\Omega_{y}R_ydy.
\end{equation}
By examining the dependence of $\mu_y,\mu_{BH}$ on the B-field \eqref{muBH}-\eqref{muy}, we can make a key observation: unlike the $\lambda$-deformed BTZ case, there is no unique way to fix the three different components of the B-field at the origin $B_{\tau\theta}^{(0)},B_{\tau y}^{(0)},B_{\theta y}^{(0)}$ solely using the thermodynamic constraints \eqref{TBH}-\eqref{muLR} combined with \eqref{omegamu-y}. One constraint is missing.
For example, in the $\Omega_{BH}=\Omega_y=0$ scenario, which compatible in describing a non-rotating background in both the angular $\theta,y$ directions, we can fix $B_{t\theta}^{(0)}$ and $B_{ty}^{(0)}$, but we lack sufficient constraints to definitively fix $B_{\theta y}^{(0)}$.

However, It was claimed in \cite{Giveon:2025cyk} that each of the B-field components at the horizon can be uniquely determined from a specific set of B-field values at the linear dilaton asymptotics. These values are required to guarantee that the long string excitation spectrum matches that of a single trace $T\bar{T}+J\bar{T}+T\bar{J}$ symmetric orbifold \eqref{spectJT}-\eqref{E_wtot TT TJ JT}.
It would be interesting to check, for a suitable deformed BTZ solution, whether the B-field components at the origin expected from "harmony with single trace $T\bar{T}+J\bar{T}+T\bar{J}$ holography" considerations, consistently satisfy the chemical potential constraints \eqref{muy}-\eqref{muBH} and \eqref{omegaBH}-\eqref{omegamu-y} dictated by the bulk thermodynamics of the winding long string emission process.
Such an investigation would extend the analysis conducted in the previous sections for the long strings emission from $\lambda$-deformed black holes  to richer $(\lambda,\epsilon_\pm)$ deformed backgrounds and single trace $T\bar{T}+J\bar{T}+T\bar{J}$ correspondence.  

\section{Discussion}
\label{sec:discussions}
In this note, we investigated the semiclassical evaporation of deformed BTZ black holes via the tunneling and WKB mechanisms, generalizing the study of \cite{Martinec:2023plo} to a class of asymptotically flat spacetimes with linear dilaton.
A direct consequence of our analysis is the necessity of a unique, non-trivial value for the B-field on the origin of $\lambda$-deformed BTZ geometry, applicable to both positive and negative deformation couplings \eqref{B-0-pos},\eqref{B-0-neg}. This uniqueness arises from the requirement of thermodynamic consistency during the emission of long strings above an energy threshold. We demonstrated that the B-field at the origin directly dictates the chemical potential in a grand canonical ensemble for a fluctuating $w$ mode. Furthermore, the resulting emission rate of a probe string universally matches the exponential change in the background black hole's entropy \eqref{Gamma2}.
As noted in \cite{Giveon:2024sgz}, these non-trivial values of $B_{\tau\theta}^{(0)}$ 
backgrounds that are both \textit{deformed} and \textit{rotating}; otherwise, vanishes. Taking the undeformed limit $\frac{R}{\sqrt{\alpha'}}\to \infty$ demonstrates consistency with the claim in \cite{Martinec:2023plo}, while taking the non-rotating limit $\rho_-\to 0$ aligns precisely with the findings for the static $\lambda$-deformed BTZ background in \cite{Chakraborty:2024ugc}.
The existence of non-zero B-field at the origin indicates the presence of a localized source at the origin, giving rise to a singular three form field\footnote{See the discussion in Section 4.2 of \cite{Chakraborty:2024mls}} $H\sim \delta^2(\rho)d\tau\wedge d\theta\wedge d\rho$. Since the B-field vanishes at the origin for both deformed and undeformed global AdS$_3$ (corresponding to $\rho_+=ir_5,\rho_-=0$), this does not contradict the sourceless, empty nature of the global vacuum geometries.
It would be interesting to investigate from the worldsheet perspective if there are gauge restrictions on the choice of the B-field at the origin, or in any choice, yields a consistent worldsheet CFT\footnote{Recently, it was argued in \cite{Apolo:2025wcl} that deformed BTZ backgrounds are compatible with single trace $T\bar{T}$ 
holography under the assumption of vanishing chemical potential. This claim, based on matching black holes \textit{torus partition function} to that of a single trace $T\bar{T}$ deformed symmetric orbifold,
 is in tension with our findings, that favors non-vanishing chemical potentials \eqref{chemical_pos},\eqref{chemical_neg}, which uniquely guarantee the right emission probability of long strings from a deformed BTZ black hole \eqref{Gamma2}. }. 

Radiation from these deformed backgrounds behaves fundamentally differently from the undeformed BTZ black hole case. In pure BTZ, emitted particle quanta eventually fall back into the black hole and joins the thermal atmosphere near its vicinity. In contrast, in positively deformed backgrounds, sufficiently light emitted particles, specifically those carrying high energy relative to their angular momentum, can detach from the black hole thermal atmosphere and escape towards the timelike infinity regime, leaving behind a smaller black hole.
 Conversely, in negatively deformed backgrounds escaping particles primarily terminate at the finite UV cutoff, never reaching true infinity.
Long string emission is similarly described as a macroscopic fragment of the thermal atmosphere that, above energy threshold "boils off" gradually the black hole. This occurs in both the undeformed BTZ \cite{Martinec:2023plo} and the deformed backgrounds, with the latter possessing a modified threshold energy \eqref{total deformed}.
In all these scenarios, the emitted long string sector reaches the boundary as plane-wave normalizable wavefunction.    
This emission process proceeds gradually until the radiation switches off as the black hole relaxes down to extremality.

As was discussed extensively in Section 4 of \cite{Martinec:2023plo},
the information paradox remains a central issue for these black hole solutions. Semiclassicaly, the emitted particles or strings are completely entangled with their Hawking partners
behind the horizon, lacking any correlation with the original microstates that initially formed the black hole. Consequently, pure quantum information that fell in during the black hole's formation seemingly fades into a thermal mixed state beyond the Page time. 
In the context of AdS/CFT correspondence, assuming the existence of an exact non-perturbative dual theory, the information originating from the initial infalling state, having scrambled within the deformed black hole, must eventually be recoverable at the boundary. The boundary theory experiences unitary time evolution as a well-defined quantum field theory, specifically for  long string states in the deformed black holes as a single trace $T\bar{T}$/  $T\bar{T}+J\bar{T}+T\bar{J}$ deformed $CFT_2s$ \cite{Chakraborty:2018kpr}. Yet, the precise bulk mechanism for this information recovery remains unclear.
 Many resolutions have been proposed in recent years. Among these there are the fuzzball scenario \cite{Bena:2022rna}, the island approach (related to the ER=EPR conjecture) \cite{Almheiri:2020cfm}, and the framework of non isometric Hilbert spaces mapping radiation to black hole microstates states\footnote{See the extensive discussions in sections 4.1-4.3 in \cite{Martinec:2023plo} and references therein.} \cite{Akers:2022qdl},\cite{DeWolfe:2023iuq}.   
 
An interesting question for future study concerns the exact dynamics and the fate of particles and long strings after they tunnel the deformed black hole horizon, and experience the exterior gravitational effective potential. In the positively deformed background, sufficiently light and energetic quanta will pass over the second barrier and reach the null infinity. Other quanta will primarily reflect off this barrier and fall back, though a small fraction may still tunnel through it.
The total outgoing asymptotic state reaching the boundary is typically evaluated via a greybody factor \cite{Hawking:1975vcx} \footnote{The study of greybody factor for stringy black hole initially treated D-brane emission in \cite{Maldacena:1996ix}, and was later on generalized to include the NS5-F1 system in \cite{Aharony:1999ti}.}.  This factor quantifies the suppression of the blackbody radiation spectrum as measured by a distant  static observer relative to the infalling near horizon emission. It would be interesting to explicitly compute this greybody factor to determine exactly how the B-field modifies the scattering amplitudes of long strings compared to those of point particles.

\vspace{1mm}

\section*{Acknowledgments}
%%%%%%%%%%%%%

I thank A. Giveon and S. Chakraborty for many helpful discussions and valuable comments on the manuscripts.
This work was supported in part by the ISF (grant number 256/22).

\appendix
\section{General Hamiltonian Path Integral Derivation}\label{appA}
%\newpage
We consider the general two-dimensional worldsheet metric of a closed string given by an ADM decomposition \cite{Arnowitt:1959ah} (for a review, see \cite{Corichi:1991qqo}):
\begin{equation} \label{WS_general}
    ds^2_{WS}=-n_{0}^{2}d\tau^{2}+h\left(d\sigma+n_{1}d\tau\right)^{2},\qquad \sigma\sim \sigma+2\pi  
\end{equation}
where $n_0$ is the \textit{lapse} function and $n_1$ is the \textit{shift} function directed along the spatial $\sigma$ direction; both are considered to be non-dynamical. The function $h$ represents the spatial part of the worldsheet metric and is considered to be a smooth functions of the worldsheet coordinates $\xi,\sigma$.
The non-linear sigma model action, describing a string propagating in a purely NS-NS background is given by the action in \eqref{I_WS}. Evaluated on the general worldsheet metric \eqref{WS_general}, the action becomes:
\begin{equation}
    I_{WS}=\frac{1}{2\pi}\int d^2\sigma\mathcal{L}_{WS}
\end{equation}
where the Lagrangian density, chosen to have a mass dimension one, is given by:
\begin{equation} \label{L_WS}
    \mathcal{L}_{WS} = -\frac{n_{0}\sqrt{h}}{2\alpha'}\left[\left(-\frac{1}{n_{0}^{2}}\left(\dot{x}^{\mu}-n_{1}x'^{\mu}\right)\left(\dot{x}^{\nu}-n_{1}x'^{\nu}\right)G_{\mu\nu}+\frac{1}{h}G_{\mu\nu}x'^{\mu}x'^{\nu}-\frac{2}{n_{0}\sqrt{h}}B_{\mu\nu}x'^{\mu}\dot{x}^{\nu}\right)+\alpha'\Phi\left(x\right)R^{\left(2\right)}\right]
\end{equation}
where overdots denote  derivatives with respect to worldsheet proper time $\xi$, and primes denote derivatives with respect to string's spatial compact coordinate $\sigma$.
The 2D scalar curvature $R^{(2)}$ can be calculated explicitly in terms of the derivatives of the spatial metric $(h,\dot{h},h',h'',\ddot{h},\dot{h}')$ and the spatial derivatives of the lapse and shift functions $(n_{0,1},n'_{0,1},n''_{0,1})$.

To transition to the Hamiltonian description, we introduce three classes of conjugate momenta:
\begin{equation}
    p_{\mu}=\frac{\partial\mathcal{L}_{WS}}{\partial{\dot{x}^\mu}},\qquad \pi_h = \frac{\partial\mathcal{L}_{WS}}{\partial{\dot{h}}},\qquad p_{\Phi} = \frac{\partial{\mathcal{L}_{WS}}}{\partial\dot{\Phi}}=\frac{\partial{\mathcal{L}_{WS}}}{\partial(\dot{x}^\mu\partial_\mu\Phi)}
\end{equation}
where $p_\mu$ is the conjugate momentum with respect to spacetime coordinate $x^{\mu}$, $\pi_h$ is the momentum conjugate to the dynamical part of the spatial slice $h$ in the worldsheet metric \eqref{WS_general}, and $p_{\Phi}$ is the momentum conjugate to the dilaton field $\Phi$. At first glance, introducing $p_\Phi$ seems peculiar since there are no explicit time derivatives of the dilaton in the original worldsheet action \eqref{I_WS}. However, following the prescription of \cite{Kuchar:1996zm} \footnote{In \cite{Kuchar:1996zm}, a model conformally related to the CGHS model was considered; its action coincides with the dilaton term in the worldsheet action when taking the limit of vanishing cosmological constant $\kappa=0$ and  no matter fields $f=0$.
The dilaton field is related to ours through the relation $y=-\frac{\Phi}{2}$ and the spatial metric through $\sigma^2=h$.
} \footnote{A review on 2D models describing dilaton coupled to gravity and theirs Hamiltonian structure appears in \cite{Grumiller:2002nm}.} , an explicit time derivative of the dilaton emerges after the ADM decomposition of the worldsheet scalar curvature $R^{(2)}$ and a time integration by parts, discarding the boundary term.

To obtain the total Hamiltonian of the worldsheet, we preform a Legendre transform on the worldsheet action \eqref{L_WS}:
\begin{equation}
    \mathcal{H}_{tot} = p_{\mu}\dot{x}^\mu+\pi_h\dot{h} -\mathcal{L}_{WS}
\end{equation}
The full Hamiltonian dynamics are described via the path integral:

\begin{equation}
    \mathcal{I}_{string} = \int\left(p_\mu\dot{x}^\mu+\pi_h \dot{h}+p_\Phi\dot{x}^\mu \partial_\mu\Phi-n_0\mathcal{H}_{string}-n_1P_{string}\right)
\end{equation}
with the \textit{Hamiltonian constraint} given by:
\begin{equation} \label{H_const_gen}
    \mathcal{H}_{string} = p_\Phi\pi_h+\frac{\alpha'}{2\sqrt{h}}\left(G^{\mu\nu}\Pi_{\mu}\Pi_{\nu}+\frac{1}{\alpha'^{2}}G_{\mu\nu}x'^{\mu}x'^{\nu}-\frac{1}{\alpha'}\mathcal{H}_\Phi+m^{2}\right)=0
\end{equation}
and the \textit{momentum constraint}:
\begin{equation}
    P_{string}=p_\mu x'^\mu+\pi_hh'+p_\Phi\partial_\mu\Phi x'^\mu=0
\end{equation}
where $m^2$ is the mass square of the string's center of mass, $\Pi_\mu$ is the spacetime kinetic momentum given in \eqref{KinPI}, and:
\begin{equation}
    \mathcal{H}_\Phi = \partial_\mu\Phi \left(x''^{\mu}-\frac{h'}{2h}x'^\mu\right)+\partial_\mu\partial_\nu\Phi x'^\mu x'^\nu.
\end{equation}
Note that the term $\mathcal{H}_\Phi$ is not contracted with any derivative of the general worldsheet metric \eqref{WS_general},  meaning it generally must appear in the Hamiltonian constraint. However, this term vanishes identically for stationary black hole solutions, because the dilaton field in these backgrounds is static and spherically symmetric (e.g. \eqref{dilaton},\eqref{dilaton_n}). Consequently, the only non-vanishing derivatives are with respect to $\rho$, which would be contracted with $\rho'$ (and $\rho''$). But, these spatial derivatives vanish identically under the spectral flow operation \eqref{momentums}.

Another affect of the dilaton on the Hamiltonian path integral, can be read from the first term in the Hamiltonian constraint \eqref{H_const_gen}. This term couples 
the dilaton and spatial metric momenta, which can be expressed explicitly in terms of the worldsheet derivatives of the dilaton field and the worldsheet metric:
\begin{equation}
    \pi_h = \sqrt{h}n_0^{-1}\partial_\mu\Phi\left(-\dot{x}^\mu+n_1x'^\mu\right),\qquad p_{\Phi} =\frac{n_0^{-1}}{2\sqrt{h}}\left(\dot{h}+2n'_1h+n_1h'\right).
\end{equation}
Substituting these into the models of interest, yields the following extra term in the Hamiltonian constraint:
\begin{equation}
    p_{\Phi}\pi_h = \frac{n_0^{-2}}{2}\left(\dot{h}+2n_1h+n_1h'\right)\partial_\rho\Phi\dot{\rho}.
\end{equation}
If we utilize the Weyl invariance of the full worldsheet theory propagating in the 10-dimensional spacetime $AdS_3\times \mathcal{N}$, we can fix the worldsheet metric to be a flat gauge \eqref{WS_metric} via the identifications:
\begin{equation}
    h\to 1,\qquad n_0^2\to \frac{1}{r_5^2},\qquad n_1\to0
\end{equation}
In this gauge, the effects of the dilaton and spatial metric fluctuations are entirely suppressed, and the Hamiltonian path integral reduces exactly to the one considered throughout this note \eqref{Istr}-\eqref{P_constr}.

\section{The Gauged-WZW $\frac{SL(2,\mathbb{R})_k\times U(1)}{U(1)}$ Model }\label{appB}
The gauged-WZW $\frac{SL(2,\mathbb{R})_k\times U(1)}{U(1)}$ model, in addition to its applications in \cite{Chakraborty:2020yka,Chakraborty:2022dgm,Apolo:2019zai} and recently via null-gauging in \cite{Massai:2025nci} to describe deformed 3D BTZ backgrounds with asymptotic linear dilaton characteristics, it was also used to describe certain charged 2D black holes solutions and scattering beyond their singularities in \cite{Giveon:2003ge}. Furthermore, it was related to the hydrodynamics of particular 2D quantum systems in \cite{Goykhman:2013oja}.  

Here, we will outline how to obtain the worldsheet action for the gauged WZW $\frac{SL(2,\mathbb{R})_k\times U(1)}{U(1)}$ action, following the derivation along the lines of \cite{Chakraborty:2020yka,Chakraborty:2022dgm,Giveon:2003ge,Giveon:2005mi,Elitzur:2002vw}.

We begin by defining a point in the product group space $(g,u)\in SL(2,\mathbb{R})_k\times U(1)$. 
Acting on the point $(g,u)$ with general $\left[U_L(1)\times U_R(1)\right]^{SL(2,\mathbb{R})_k}\times U_L(1)\times U_R(1)$ transformation, where $\left[U_L(1)\times U_R(1)\right]^{SL(2,\mathbb{R})_k}$ are the transformations associated with the left and right sectors of the azimuthal part of the $SL(2,
\mathbb{R})_k$ manifold $g$, and $U_L(1)\times U_R(1)$ are the left and right sectors of the compact coordinate $\tilde{x}\sim \tilde{x} +2\pi R$, gives:
\begin{equation} \label{transforms}
    (g,u_L,u_R)\to (e^{r_1\sigma_3}ge^{q_1\sigma_3},\tilde{x}_L+r_2,\tilde{x}_R+q_2)
\end{equation}
The right moving transformation can be described by a vector with length $|\vec{q}|=q$ projected onto its components with respect to an angle $\chi$, such that:
\begin{equation} 
    \vec{q}=\left(\begin{array}{c}
q_{1}\\
q_{2}
\end{array}\right)=q\Vec{V},\qquad \vec{V}= \left(\begin{array}{c}
\cos\left(\chi\right)\\
\sin\left(\chi\right)
\end{array}\right).
\end{equation}
The WZW model of the product manifold $SL(2,\mathbb{R})_k\times U(1)$ is described by the WZW action given in \eqref{WZW} by replacing $g\to G$, where $G$ is given by:
\begin{equation}
     G=\left(\begin{array}{cc}
g & 0\\
0 & e^{\sqrt{2}\tilde{x}}
\end{array}\right).
\end{equation}
The substitution of $G$ yields the decomposition:
\begin{equation} \label{WZW_dec}
    I=I_{\text{WZW}}[g]+\underbrace{\frac{k}{2\pi}\int_{\Sigma} d^2z\partial \tilde{x}\bar{\partial} \tilde{x}}_{I[\tilde{x}]}
\end{equation}
$G$ transforms according to \eqref{transforms} as follows:
\begin{equation}
    G\to e^{T_L}Ge^{T_R}
\end{equation}
with
\begin{equation}
    T_{L}=\left(\begin{array}{cc}
r_{1}\sigma_{3} & 0\\
0 & r_{2}
\end{array}\right),\quad T_{R}=\left(\begin{array}{cc}
q_{1}\sigma_{3} & 0\\
0 & q_{2}
\end{array}\right).
\end{equation}
To ensure anomaly free gauging, one requires that:
\begin{equation} \label{Req}
    \text{Tr}(T_L^2)=\text{Tr}(T_R^2)
\end{equation}
As consequence of the requirement \eqref{Req},  the vector representing the left handed transformations must have the same norm as the right handed vector: $\sqrt{r_1^2+r_2^2}=|\vec q|=q$. In general, it is related to the right handed vector by a rotation matrix $\mathcal{R}$ with an angle $\psi$:
\begin{equation}
    \mathcal{R}\vec{q}=q\left(\begin{array}{c}
\cos\left(\chi-\psi\right)\\
\sin\left(\chi-\psi\right)
\end{array}\right),\quad \mathcal{R}=\left(\begin{array}{cc}
\cos\left(\psi\right) & \sin\left(\psi\right)\\
-\sin\left(\psi\right) & \cos\left(\psi\right)
\end{array}\right)
\end{equation}
Applying the transformation \eqref{transforms} to the decomposed worldsheet action \label{WZW_dec} and treating the length $q$  as a dynamical field $q\equiv \hat{q}(\xi,\sigma)$ gives the following action:
\begin{equation} \label{decpm}
  I=I_{\text{WZW}}[e^{\hat{r}_1\sigma_3}ge^{\hat{q}_1\sigma_3}]+I[\tilde{x}_L+\hat{r}_2,\tilde{x}_R+\hat{q}_2] -\frac{k}{2\pi} \int_{\Sigma} d^2z(\mathcal{R}\partial\vec{\hat{q}}-\partial\mathcal{R}\vec{\hat{q}})^t(\mathcal{R}\bar{\partial}\vec{\hat{q}}-\bar{\partial}\mathcal{R}\vec{\hat{q}})
\end{equation}
Using the Polyakov-Wiegmann identity:
\begin{multline}
I[ABC]=I[A]+I[B]+I[C] -\\ \frac{k}{2\pi}\int d^{2}z \left(A^{-1} \partial A B^{-1}\bar{\partial} B+B^{-1}\partial B C^{-1}\bar{\partial}C+A^{-1}\partial A B C^{-1}\bar{\partial}C B^{-1}\right)
\end{multline}
the action \eqref{decpm} can be  brought into the form:
\begin{equation}
 I=I_{\text{WZW}}[g]+I[\tilde{x}]+\frac{k}{2\pi}\int _{\Sigma}d^2z (A\bar{\textbf{J}}+\bar{A}\textbf{J}+2A\bar{A}(\mathcal{R}\vec V)^tM\vec{V}).
\end{equation}
The left and right currents $\bar{\boldsymbol{{J}}},\boldsymbol{J}$ are given by:
\begin{equation} \label{currents}
\begin{aligned}
\boldsymbol{J} =& \left(\text{Tr}[\partial g g^{-1} \sigma_3],2\partial \tilde{x}\right)\mathcal{R}\vec{V}=\cos\left(\chi-\psi\right)\text{Tr}[\partial g g^{-1} \sigma_3]+2\sin\left(\chi-\psi\right)\partial \tilde{x} \\
\bar{\boldsymbol{J}} = & \left(\text{Tr}[g^{-1} \bar{\partial} g  \sigma_3],2\bar{\partial} \tilde{x}\right)\vec{V} = \cos\left(\chi\right)\text{Tr}[g^{-1} \bar{\partial} g  \sigma_3]-2\sin\left(\chi\right)\bar{\partial} \tilde{x}
\end{aligned}
\end{equation}
while the auxiliary fields $A,\bar{A}$ are given by:
\begin{equation}    A=\partial\hat{q},\quad\bar{A}=-\bar{\partial}\hat{q}.
\end{equation}
and the matrix $M$ is given by:
\begin{equation}
  M=\left(\begin{array}{cc}
\frac{1}{2}Tr\left(g^{-1}\sigma_{3}g\sigma_{3}\right) & 0\\
0 & 1
\end{array}\right)+\mathcal{R}.
\end{equation}
Gauging out the auxiliary left and right $U(1)$ gauge fields $A,\bar{A}$ from $SL(2,\mathbb{R})_k\times U(1)$ WZW model, to form eventually a gauged-WZW $\frac{SL(2,\mathbb{R})_k\times U(1)}{U(1)}$ model, governed by the  worldsheet action:
\begin{equation} \label{gWZW}  I=I_{\text{WZW}}\left[g\right]+I\left[\tilde{x}\right]-\frac{k}{4\pi}\int_{\Sigma} d^{2}z\frac{\bar{\boldsymbol{J}}\boldsymbol{J}}{\left(\mathcal{R}\vec{V}\right)^{t}M\vec{V}}
\end{equation}
accompanied by the deformed dilaton field:
\begin{equation}
    \Phi_d = \Phi-\frac{1}{2}\log\left(\left(\mathcal{R}\vec{V}\right)^{t}M\vec{V}\right).
\end{equation}
To evaluate the action, one should specify a particular parametrization for $g\in SL(2,\mathbb{R})$. In general, the $SL(2,\mathbb{R})$ element can be written in terms of Euler angles $r,\tilde{u},\tilde{v}$
\begin{equation}
g = e^{\tilde{u}\sigma_3}e^{r\sigma_1}e^{\tilde{v}\sigma_3}.
\end{equation}
We will not provide here the explicit form of the worldsheet action under this general parametrization here. Instead, we specify the exact parametrization, group embedding identifications, and coordinate transformations that directly lead to the $\lambda$-deformed solution \eqref{defbtz}-\eqref{dilaton}.
We consider the BTZ parametrization \eqref{gBTZ}:
\begin{equation} \label{BTZparam}
 \tilde{u}=\frac{\rho_{+}-\rho_{-}}{2r_{5}}\left(\frac{\tilde{\tau}}{r_{5}}+\tilde{\theta}\right),\qquad \tilde{v} = -\frac{\rho_{+}+\rho_{-}}{2r_{5}}\left(\frac{\tilde{\tau}}{r_{5}}-\tilde{\theta}\right)
\end{equation}
together with the radial relation \eqref{redf}. Here we use tilde notations to distinguish the coordinates of the undefrormed theory from the deformed one. Under this parametrization, the worldsheet action \eqref{gWZW} corresponds to a 4D black string solution in $(\rho,\tilde{\tau},\tilde{\theta},\tilde{x})$ coordinates. To obtain the deformed 3D background, we would fix the $U(1)$ gauge symmetry by setting $\tilde{x}=0$.

Alternatively, it was shown in \cite{Chakraborty:2020yka,Chakraborty:2022dgm} that similar backgrounds can be obtained by gauge fixing the symmetry on the group manifold $g\in SL(2,\mathbb{R})$  instead, by setting $\tilde{u}=-\tilde{v}=\frac{\tilde{y}}{2}$. In this case, $g$ represents a slice of the BTZ manifold amounts to a Rindler $AdS_2$ space.     

Using the BTZ parametrization \eqref{BTZparam} alongside the identification the group embedding angles $\chi,\psi$ with the physical black hole parameters:
\begin{equation} \label{psichi}
\cos\left(\psi\right)=\sqrt{\frac{1+\frac{\rho_{+}^{2}}{R^{2}}}{1+\frac{\rho_{-}^{2}}{R^{2}}}}-1,\quad\cos\left(2\chi-\psi\right)=1-\sqrt{\frac{1+\frac{\rho_{-}^{2}}{R^{2}}}{1+\frac{\rho_{+}^{2}}{R^{2}}}}
\end{equation}
and applying the coordinate transformations:

\begin{equation} \label{tildeton}
\begin{aligned}
&  \tilde{\tau}={\frac{1}{\rho_{+}^{2}-\rho_{-}^{2}}} \left[\left(\frac{\rho_{+}^{2}}{\sqrt{1+\frac{\rho_{+}^{2}}{R^{2}}}}-\frac{\rho_{-}^{2}}{\sqrt{1+\frac{\rho_{-}^{2}}{R^{2}}}}\right)\tau-r_{5}\rho_{-}\rho_{+}\left(\frac{1}{\sqrt{1+\frac{\rho_{+}^{2}}{R^{2}}}}-\frac{1}{\sqrt{1+\frac{\rho_{-}^{2}}{R^{2}}}}\right)\theta\right] \\
&  \tilde{\theta} = \frac{1}{\rho_{+}^{2}-\rho_{-}^{2}}\left[\frac{\rho_{-}\rho_{+}}{r_{5}}\left(\frac{1}{\sqrt{1+\frac{\rho_{+}^{2}}{R^{2}}}}-\frac{1}{\sqrt{1+\frac{\rho_{-}^{2}}{R^{2}}}}\right)\tau+\left(\frac{\rho_{+}^{2}}{\sqrt{1+\frac{\rho_{-}^{2}}{R^{2}}}}-\frac{\rho_{-}^{2}}{\sqrt{1+\frac{\rho_{+}^{2}}{R^{2}}}}\right)\theta\right]
\end{aligned}
\end{equation}
yields the desired deformed BTZ solution \eqref{defbtz}-\eqref{dilaton}.
The negatively deformed background \eqref{defbtz_n}-\eqref{dilaton_n}, is obtained by \eqref{psichi},\eqref{tildeton} under the transformation $R^2\to -R^2$.
If we assume that the underlying BTZ background has a vanishing B-field at the origin, an assumption supported in \cite{Martinec:2023plo,Giveon:2024sgz}, the worldsheet coset construction above,naturally gives rise to a preferred value for $B_{\tau\theta}(\rho=0)=B^{(0)}_{\tau\theta}$. This uniquely derived constant matches exactly with \eqref{B^0_p} for positive deformations and \eqref{B^0_n} for negative deformations. These are precisely the identical values guaranteed by the thermodynamic consistency of the long string emission process, as well as by the algebraic requirements of the single-trace $T\bar{T}$ energy formula.

\bibliographystyle{JHEP}      

\bibliography{References}

@article{Giveon:1998ns,
    author = "Giveon, Amit and Kutasov, D. and Seiberg, Nathan",
    title = "{Comments on string theory on AdS(3)}",
    eprint = "hep-th/9806194",
    archivePrefix = "arXiv",
    reportNumber = "EFI-98-22, RI-4-98, IASSNS-HEP-98-52",
    doi = "10.4310/ATMP.1998.v2.n4.a3",
    journal = "Adv. Theor. Math. Phys.",
    volume = "2",
    pages = "733--782",
    year = "1998"
}

@article{Wu:2006pz,
    author = "Wu, Shuang-Qing and Jiang, Qing-Quan",
    title = "{Remarks on Hawking radiation as tunneling from the BTZ black holes}",
    eprint = "hep-th/0602033",
    archivePrefix = "arXiv",
    doi = "10.1088/1126-6708/2006/03/079",
    journal = "JHEP",
    volume = "03",
    pages = "079",
    year = "2006"
}

@article{Martinec:2023plo,
    author = "Martinec, Emil J.",
    title = "{The holar wind}",
    eprint = "2303.00234",
    archivePrefix = "arXiv",
    primaryClass = "hep-th",
    doi = "10.1007/JHEP07(2023)113",
    journal = "JHEP",
    volume = "07",
    pages = "113",
    year = "2023"
}

@article{Li:2006rg,
    author = "Li, Hui-Ling and Yang, Shu-Zheng and Jiang, Qing-Quan and Qi, De-Jiang",
    title = "{Charged particle's tunneling radiation from the charged BTZ black hole}",
    doi = "10.1016/j.physletb.2006.08.033",
    journal = "Phys. Lett. B",
    volume = "641",
    pages = "139--144",
    year = "2006"
}

@article{Parikh:1999mf,
    author = "Parikh, Maulik K. and Wilczek, Frank",
    title = "{Hawking radiation as tunneling}",
    eprint = "hep-th/9907001",
    archivePrefix = "arXiv",
    reportNumber = "PUPT-1775, SPIN-1998-12, IASSNS-HEP-98-22",
    doi = "10.1103/PhysRevLett.85.5042",
    journal = "Phys. Rev. Lett.",
    volume = "85",
    pages = "5042--5045",
    year = "2000"
}

@article{Vanzo:2011wq,
    author = "Vanzo, L. and Acquaviva, G. and Di Criscienzo, R.",
    title = "{Tunnelling Methods and Hawking's radiation: achievements and prospects}",
    eprint = "1106.4153",
    archivePrefix = "arXiv",
    primaryClass = "gr-qc",
    doi = "10.1088/0264-9381/28/18/183001",
    journal = "Class. Quant. Grav.",
    volume = "28",
    pages = "183001",
    year = "2011"
}

@article{Chakraborty:2023zdd,
    author = "Chakraborty, Soumangsu and Giveon, Amit and Kutasov, David",
    title = "{Momentum in Single-trace $T\bar T$ Holography}",
    eprint = "2304.09212",
    archivePrefix = "arXiv",
    primaryClass = "hep-th",
    doi = "10.1016/j.nuclphysb.2023.116405",
    journal = "Nucl. Phys. B",
    volume = "998",
    pages = "116405",
    year = "2024"
}

@article{Giveon:2024sgz,
    author = "Giveon, Amit and Vainshtein, Daniel",
    title = "{To B or not to be in single-trace TT{\textasciimacron} holography}",
    eprint = "2408.03022",
    archivePrefix = "arXiv",
    primaryClass = "hep-th",
    doi = "10.1016/j.nuclphysb.2025.116905",
    journal = "Nucl. Phys. B",
    volume = "1015",
    pages = "116905",
    year = "2025"
}

@article{Chakraborty:2024ugc,
    author = "Chakraborty, Soumangsu and Giveon, Amit and Hashimoto, Akikazu",
    title = "{On string theory on deformed BTZ and $ \textrm{T}\overline{\textrm{T}} $}",
    eprint = "2402.05776",
    archivePrefix = "arXiv",
    primaryClass = "hep-th",
    doi = "10.1007/JHEP04(2024)134",
    journal = "JHEP",
    volume = "04",
    pages = "134",
    year = "2024"
}

@article{Chakraborty:2020swe,
    author = "Chakraborty, Soumangsu and Giveon, Amit and Kutasov, David",
    title = "{$ T\overline{T} $, black holes and negative strings}",
    eprint = "2006.13249",
    archivePrefix = "arXiv",
    primaryClass = "hep-th",
    doi = "10.1007/JHEP09(2020)057",
    journal = "JHEP",
    volume = "09",
    pages = "057",
    year = "2020"
}

@article{Dijkgraaf:2016lym,
    author = "Dijkgraaf, Robbert and Heidenreich, Ben and Jefferson, Patrick and Vafa, Cumrun",
    title = "{Negative Branes, Supergroups and the Signature of Spacetime}",
    eprint = "1603.05665",
    archivePrefix = "arXiv",
    primaryClass = "hep-th",
    doi = "10.1007/JHEP02(2018)050",
    journal = "JHEP",
    volume = "02",
    pages = "050",
    year = "2018"
}

@article{Hemming:2000as,
    author = "Hemming, Samuli and Keski-Vakkuri, Esko",
    title = "{Hawking radiation from AdS black holes}",
    eprint = "gr-qc/0005115",
    archivePrefix = "arXiv",
    reportNumber = "HIP-2000-25-TH",
    doi = "10.1103/PhysRevD.64.044006",
    journal = "Phys. Rev. D",
    volume = "64",
    pages = "044006",
    year = "2001"
}

@article{Maldacena:2000hw,
    author = "Maldacena, Juan Martin and Ooguri, Hirosi",
    title = "{Strings in AdS(3) and SL(2,R) WZW model 1.: The Spectrum}",
    eprint = "hep-th/0001053",
    archivePrefix = "arXiv",
    reportNumber = "CALT-68-2245, CITUSC-99-010, HUTP-99-A027, LBNL-44375, UCB-PTH-99-48, LBL-44375",
    doi = "10.1063/1.1377273",
    journal = "J. Math. Phys.",
    volume = "42",
    pages = "2929--2960",
    year = "2001"
}

@article{Chakraborty:2020yka,
    author = "Chakraborty, Soumangsu",
    title = "{$ \frac{\mathrm{SL}\left(2,\mathrm{\mathbb{R}}\right)\times \mathrm{U}(1)}{\mathrm{U}(1)} $ CFT, NS5+F1 system and single trace $ T\overline{T} $}",
    eprint = "2012.03995",
    archivePrefix = "arXiv",
    primaryClass = "hep-th",
    doi = "10.1007/JHEP03(2021)113",
    journal = "JHEP",
    volume = "03",
    pages = "113",
    year = "2021"
}

@article{Giveon:2025cyk,
    author = "Giveon, Amit and Vainshtein, Daniel",
    title = "{On string theory on deformed BTZ and $ T\overline{T}+J\overline{T}+T\overline{J} $}",
    eprint = "2507.22595",
    archivePrefix = "arXiv",
    primaryClass = "hep-th",
    doi = "10.1007/JHEP02(2026)164",
    journal = "JHEP",
    volume = "02",
    pages = "164",
    year = "2026"
}

@article{Kutasov:1999xu,
    author = "Kutasov, David and Seiberg, Nathan",
    title = "{More comments on string theory on AdS(3)}",
    eprint = "hep-th/9903219",
    archivePrefix = "arXiv",
    reportNumber = "EFI-99-8, IASSNS-HEP-99-30",
    doi = "10.1088/1126-6708/1999/04/008",
    journal = "JHEP",
    volume = "04",
    pages = "008",
    year = "1999"
}

@article{Argurio:2000tb,
    author = "Argurio, Riccardo and Giveon, Amit and Shomer, Assaf",
    title = "{Superstrings on AdS(3) and symmetric products}",
    eprint = "hep-th/0009242",
    archivePrefix = "arXiv",
    reportNumber = "RI-3-00",
    doi = "10.1088/1126-6708/2000/12/003",
    journal = "JHEP",
    volume = "12",
    pages = "003",
    year = "2000"
}

@article{Giveon:2005mi,
    author = "Giveon, A. and Kutasov, D. and Rabinovici, E. and Sever, A.",
    title = "{Phases of quantum gravity in AdS(3) and linear dilaton backgrounds}",
    eprint = "hep-th/0503121",
    archivePrefix = "arXiv",
    doi = "10.1016/j.nuclphysb.2005.04.015",
    journal = "Nucl. Phys. B",
    volume = "719",
    pages = "3--34",
    year = "2005"
}

@article{Asrat:2017tzd,
    author = "Asrat, Meseret and Giveon, Amit and Itzhaki, Nissan and Kutasov, David",
    title = "{Holography Beyond AdS}",
    eprint = "1711.02690",
    archivePrefix = "arXiv",
    primaryClass = "hep-th",
    doi = "10.1016/j.nuclphysb.2018.05.005",
    journal = "Nucl. Phys. B",
    volume = "932",
    pages = "241--253",
    year = "2018"
}

@article{Giveon:2017myj,
    author = "Giveon, Amit and Itzhaki, Nissan and Kutasov, David",
    title = "{A solvable irrelevant deformation of AdS$_{3}$/CFT$_{2}$}",
    eprint = "1707.05800",
    archivePrefix = "arXiv",
    primaryClass = "hep-th",
    doi = "10.1007/JHEP12(2017)155",
    journal = "JHEP",
    volume = "12",
    pages = "155",
    year = "2017"
}

@article{Giveon:2017nie,
    author = "Giveon, Amit and Itzhaki, Nissan and Kutasov, David",
    title = "{$ \mathrm{T}\overline{\mathrm{T}} $ and LST}",
    eprint = "1701.05576",
    archivePrefix = "arXiv",
    primaryClass = "hep-th",
    doi = "10.1007/JHEP07(2017)122",
    journal = "JHEP",
    volume = "07",
    pages = "122",
    year = "2017"
}

@article{He:2025ppz,
    author = "He, Song and Li, Yi and Ouyang, Hao and Sun, Yuan",
    title = "{$T\overline{T}$ deformation: Introduction and some recent advances}",
    eprint = "2503.09997",
    archivePrefix = "arXiv",
    primaryClass = "hep-th",
    doi = "10.1007/s11433-025-2708-2",
    journal = "Sci. China Phys. Mech. Astron.",
    volume = "68",
    number = "10",
    pages = "101001",
    year = "2025"
}

@article{Smirnov:2016lqw,
    author = "Smirnov, F. A. and Zamolodchikov, A. B.",
    title = "{On space of integrable quantum field theories}",
    eprint = "1608.05499",
    archivePrefix = "arXiv",
    primaryClass = "hep-th",
    doi = "10.1016/j.nuclphysb.2016.12.014",
    journal = "Nucl. Phys. B",
    volume = "915",
    pages = "363--383",
    year = "2017"
}

@article{Cavaglia:2016oda,
    author = "Cavagli{\`a}, Andrea and Negro, Stefano and Sz{\'e}cs{\'e}nyi, Istv{\'a}n M. and Tateo, Roberto",
    title = "{$T \bar{T}$-deformed 2D Quantum Field Theories}",
    eprint = "1608.05534",
    archivePrefix = "arXiv",
    primaryClass = "hep-th",
    doi = "10.1007/JHEP10(2016)112",
    journal = "JHEP",
    volume = "10",
    pages = "112",
    year = "2016"
}

@article{Giribet:2017imm,
    author = "Giribet, Gaston",
    title = "{$T\bar{T}$-deformations, AdS/CFT and correlation functions}",
    eprint = "1711.02716",
    archivePrefix = "arXiv",
    primaryClass = "hep-th",
    doi = "10.1007/JHEP02(2018)114",
    journal = "JHEP",
    volume = "02",
    pages = "114",
    year = "2018"
}

@article{Chen:2018eqk,
    author = "Chen, Bin and Chen, Lin and Hao, Peng-Xiang",
    title = "{Entanglement entropy in $T\overline{T}$-deformed CFT}",
    eprint = "1807.08293",
    archivePrefix = "arXiv",
    primaryClass = "hep-th",
    doi = "10.1103/PhysRevD.98.086025",
    journal = "Phys. Rev. D",
    volume = "98",
    number = "8",
    pages = "086025",
    year = "2018"
}

@article{Chakraborty:2020udr,
    author = "Chakraborty, Soumangsu and Hashimoto, Akikazu",
    title = "{Entanglement entropy for $ \mathrm{T}\overline{\mathrm{T}} $, $ \mathrm{J}\overline{\mathrm{T}} $, $ \mathrm{T}\overline{\mathrm{J}} $ deformed holographic CFT}",
    eprint = "2010.15759",
    archivePrefix = "arXiv",
    primaryClass = "hep-th",
    doi = "10.1007/JHEP02(2021)096",
    journal = "JHEP",
    volume = "02",
    pages = "096",
    year = "2021"
}

@article{Chakraborty:2018kpr,
    author = "Chakraborty, Soumangsu and Giveon, Amit and Itzhaki, Nissan and Kutasov, David",
    title = "{Entanglement beyond AdS}",
    eprint = "1805.06286",
    archivePrefix = "arXiv",
    primaryClass = "hep-th",
    doi = "10.1016/j.nuclphysb.2018.08.011",
    journal = "Nucl. Phys. B",
    volume = "935",
    pages = "290--309",
    year = "2018"
}

@article{Apolo:2019zai,
    author = "Apolo, Luis and Detournay, Stephane and Song, Wei",
    title = "{TsT, $T\bar{T}$ and black strings}",
    eprint = "1911.12359",
    archivePrefix = "arXiv",
    primaryClass = "hep-th",
    doi = "10.1007/JHEP06(2020)109",
    journal = "JHEP",
    volume = "06",
    pages = "109",
    year = "2020"
}

@article{Chakraborty:2023mzc,
    author = "Chakraborty, Soumangsu and Giveon, Amit and Kutasov, David",
    title = "{Comments on single-trace $ T\overline{T} $ holography}",
    eprint = "2303.12422",
    archivePrefix = "arXiv",
    primaryClass = "hep-th",
    doi = "10.1007/JHEP06(2023)018",
    journal = "JHEP",
    volume = "06",
    pages = "018",
    year = "2023"
}

@article{Chang:2023kkq,
    author = "Chang, Chih-Kai and Ferko, Christian and Sethi, Savdeep",
    title = "{Holography and irrelevant operators}",
    eprint = "2302.03041",
    archivePrefix = "arXiv",
    primaryClass = "hep-th",
    reportNumber = "EFI-21-8",
    doi = "10.1103/PhysRevD.107.126021",
    journal = "Phys. Rev. D",
    volume = "107",
    number = "12",
    pages = "126021",
    year = "2023"
}

@article{Hemming:2002kd,
    author = "Hemming, Samuli and Keski-Vakkuri, Esko and Kraus, Per",
    title = "{Strings in the extended BTZ space-time}",
    eprint = "hep-th/0208003",
    archivePrefix = "arXiv",
    reportNumber = "HIP-2002-30-TH, UCLA-02-TEP-21",
    doi = "10.1088/1126-6708/2002/10/006",
    journal = "JHEP",
    volume = "10",
    pages = "006",
    year = "2002"
}

@article{Troost:2002wk,
    author = "Troost, Jan",
    title = "{Winding strings and AdS(3) black holes}",
    eprint = "hep-th/0206118",
    archivePrefix = "arXiv",
    reportNumber = "MIT-CTP-3277",
    doi = "10.1088/1126-6708/2002/09/041",
    journal = "JHEP",
    volume = "09",
    pages = "041",
    year = "2002"
}

@article{Chakraborty:2019mdf,
    author = "Chakraborty, Soumangsu and Giveon, Amit and Kutasov, David",
    title = "{$T\bar{T}$, $J\bar{T}$, $T\bar{J}$ and String Theory}",
    eprint = "1905.00051",
    archivePrefix = "arXiv",
    primaryClass = "hep-th",
    doi = "10.1088/1751-8121/ab3710",
    journal = "J. Phys. A",
    volume = "52",
    number = "38",
    pages = "384003",
    year = "2019"
}

@article{Apolo:2021wcn,
    author = "Apolo, Luis and Song, Wei",
    title = "{TsT, black holes, and $ T\overline{T} $ + $ J\overline{T} $ + $ T\overline{J} $}",
    eprint = "2111.02243",
    archivePrefix = "arXiv",
    primaryClass = "hep-th",
    doi = "10.1007/JHEP04(2022)177",
    journal = "JHEP",
    volume = "04",
    pages = "177",
    year = "2022"
}

@article{Guica:2017lia,
    author = "Guica, Monica",
    title = "{An integrable Lorentz-breaking deformation of two-dimensional CFTs}",
    eprint = "1710.08415",
    archivePrefix = "arXiv",
    primaryClass = "hep-th",
    doi = "10.21468/SciPostPhys.5.5.048",
    journal = "SciPost Phys.",
    volume = "5",
    number = "5",
    pages = "048",
    year = "2018"
}

@article{Chakraborty:2018vja,
    author = "Chakraborty, Soumangsu and Giveon, Amit and Kutasov, David",
    title = "{$ J\overline{T} $ deformed CFT$_{2}$ and string theory}",
    eprint = "1806.09667",
    archivePrefix = "arXiv",
    primaryClass = "hep-th",
    doi = "10.1007/JHEP10(2018)057",
    journal = "JHEP",
    volume = "10",
    pages = "057",
    year = "2018"
}

@article{LeFloch:2019rut,
    author = "Le Floch, Bruno and Mezei, M{\'a}rk",
    title = "{Solving a family of $T\bar{T}$-like theories}",
    eprint = "1903.07606",
    archivePrefix = "arXiv",
    primaryClass = "hep-th",
    month = "3",
    year = "2019"
}

@article{Giveon:2019fgr,
    author = "Giveon, Amit",
    title = "{Comments on $T\bar T$, $J\bar{T}$ and String Theory}",
    eprint = "1903.06883",
    archivePrefix = "arXiv",
    primaryClass = "hep-th",
    month = "3",
    year = "2019"
}

@article{Katoch:2022hdf,
    author = "Katoch, Gaurav and Mitra, Swejyoti and Roy, Shubho R.",
    title = "{Holographic complexity of LST and single trace $ T\overline{T} $, $ J\overline{T} $ and $ T\overline{J} $ deformations}",
    eprint = "2208.02314",
    archivePrefix = "arXiv",
    primaryClass = "hep-th",
    doi = "10.1007/JHEP10(2022)143",
    journal = "JHEP",
    volume = "10",
    pages = "143",
    year = "2022"
}

@article{Arnowitt:1959ah,
    author = "Arnowitt, Richard L. and Deser, Stanley and Misner, Charles W.",
    title = "{Dynamical Structure and Definition of Energy in General Relativity}",
    doi = "10.1103/PhysRev.116.1322",
    journal = "Phys. Rev.",
    volume = "116",
    pages = "1322--1330",
    year = "1959"
}

@article{Corichi:1991qqo,
    author = "Corichi, Alejandro and N{\'u}{\~n}ez, Dario",
    title = "{Introduction to the ADM formalism}",
    eprint = "2210.10103",
    archivePrefix = "arXiv",
    primaryClass = "gr-qc",
    journal = "Rev. Mex. Fis.",
    volume = "37",
    pages = "720--747",
    year = "1991"
}

@article{Farina:1993xw,
    author = "Farina, C. and Gamboa, J. and Segui-Santonja, Antonio J.",
    title = "{Motion and trajectories of particles around three-dimensional black holes}",
    eprint = "gr-qc/9303005",
    archivePrefix = "arXiv",
    reportNumber = "IPNO-TH-93-06, DFTUZ-93-01",
    doi = "10.1088/0264-9381/10/11/001",
    journal = "Class. Quant. Grav.",
    volume = "10",
    pages = "L193--L200",
    year = "1993"
}

@article{Kuchar:1996zm,
    author = "Kuchar, Karel V. and Romano, Joseph D. and Varadarajan, Madhavan",
    title = "{Dirac constraint quantization of a dilatonic model of gravitational collapse}",
    eprint = "gr-qc/9608011",
    archivePrefix = "arXiv",
    doi = "10.1103/PhysRevD.55.795",
    journal = "Phys. Rev. D",
    volume = "55",
    pages = "795--808",
    year = "1997"
}

@article{Grumiller:2002nm,
    author = "Grumiller, D. and Kummer, W. and Vassilevich, D. V.",
    title = "{Dilaton gravity in two-dimensions}",
    eprint = "hep-th/0204253",
    archivePrefix = "arXiv",
    reportNumber = "TUW-02-01",
    doi = "10.1016/S0370-1573(02)00267-3",
    journal = "Phys. Rept.",
    volume = "369",
    pages = "327--430",
    year = "2002"
}

@article{Cruz:1994ir,
    author = "Cruz, Norman and Martinez, Cristian and Pena, Leda",
    title = "{Geodesic structure of the (2+1) black hole}",
    eprint = "gr-qc/9401025",
    archivePrefix = "arXiv",
    doi = "10.1088/0264-9381/11/11/014",
    journal = "Class. Quant. Grav.",
    volume = "11",
    pages = "2731--2740",
    year = "1994"
}

@article{Compere:2007vx,
    author = "Compere, Geoffrey",
    title = "{Note on the First Law with p-form potentials}",
    eprint = "hep-th/0703004",
    archivePrefix = "arXiv",
    reportNumber = "ULB-TH-07-10",
    doi = "10.1103/PhysRevD.75.124020",
    journal = "Phys. Rev. D",
    volume = "75",
    pages = "124020",
    year = "2007"
}

@article{Giveon:2003ge,
    author = "Giveon, Amit and Rabinovici, Eliezer and Sever, Amit",
    title = "{Beyond the singularity of the 2-D charged black hole}",
    eprint = "hep-th/0305140",
    archivePrefix = "arXiv",
    reportNumber = "RI-05-03",
    doi = "10.1088/1126-6708/2003/07/055",
    journal = "JHEP",
    volume = "07",
    pages = "055",
    year = "2003"
}

@article{Goykhman:2013oja,
    author = "Goykhman, Mikhail and Parnachev, Andrei",
    title = "{Stringy holography at finite density}",
    eprint = "1304.4496",
    archivePrefix = "arXiv",
    primaryClass = "hep-th",
    doi = "10.1016/j.nuclphysb.2013.05.011",
    journal = "Nucl. Phys. B",
    volume = "874",
    pages = "115--146",
    year = "2013"
}

@article{Elitzur:2002vw,
    author = "Elitzur, Shmuel and Giveon, Amit and Rabinovici, Eliezer",
    title = "{Removing singularities}",
    eprint = "hep-th/0212242",
    archivePrefix = "arXiv",
    reportNumber = "RI-05-02",
    doi = "10.1088/1126-6708/2003/01/017",
    journal = "JHEP",
    volume = "01",
    pages = "017",
    year = "2003"
}

@article{Chakraborty:2024mls,
    author = "Chakraborty, Soumangsu and Giveon, Amit and Hashimoto, Akikazu",
    title = "{Thermal partition function of $ {J}_3{\overline{J}}_3 $ deformed AdS$_{3}$}",
    eprint = "2403.03979",
    archivePrefix = "arXiv",
    primaryClass = "hep-th",
    doi = "10.1007/JHEP07(2024)141",
    journal = "JHEP",
    volume = "07",
    pages = "141",
    year = "2024"
}

@article{Apolo:2025wcl,
    author = "Apolo, Luis",
    title = "{The on-shell action of supergravity {\&} the B-side of TsT and single-trace $ T\overline{T} $}",
    eprint = "2508.19246",
    archivePrefix = "arXiv",
    primaryClass = "hep-th",
    doi = "10.1007/JHEP03(2026)098",
    journal = "JHEP",
    volume = "03",
    pages = "098",
    year = "2026"
}

@article{Bena:2022rna,
    author = "Bena, Iosif and Martinec, Emil J. and Mathur, Samir D. and Warner, Nicholas P.",
    title = "{Fuzzballs and Microstate Geometries: Black-Hole Structure in String Theory}",
    eprint = "2204.13113",
    archivePrefix = "arXiv",
    primaryClass = "hep-th",
    month = "4",
    year = "2022"
}

@article{Almheiri:2020cfm,
    author = "Almheiri, Ahmed and Hartman, Thomas and Maldacena, Juan and Shaghoulian, Edgar and Tajdini, Amirhossein",
    title = "{The entropy of Hawking radiation}",
    eprint = "2006.06872",
    archivePrefix = "arXiv",
    primaryClass = "hep-th",
    doi = "10.1103/RevModPhys.93.035002",
    journal = "Rev. Mod. Phys.",
    volume = "93",
    number = "3",
    pages = "035002",
    year = "2021"
}

@article{Akers:2022qdl,
    author = "Akers, Chris and Engelhardt, Netta and Harlow, Daniel and Penington, Geoff and Vardhan, Shreya",
    title = "{The black hole interior from non-isometric codes and complexity}",
    eprint = "2207.06536",
    archivePrefix = "arXiv",
    primaryClass = "hep-th",
    doi = "10.1007/JHEP06(2024)155",
    journal = "JHEP",
    volume = "06",
    pages = "155",
    year = "2024"
}

@article{DeWolfe:2023iuq,
    author = "DeWolfe, Oliver and Higginbotham, Kenneth",
    title = "{Non-isometric codes for the black hole interior from fundamental and effective dynamics}",
    eprint = "2304.12345",
    archivePrefix = "arXiv",
    primaryClass = "hep-th",
    doi = "10.1007/JHEP09(2023)068",
    journal = "JHEP",
    volume = "09",
    pages = "068",
    year = "2023"
}

@article{Maldacena:1996ix,
    author = "Maldacena, Juan Martin and Strominger, Andrew",
    title = "{Black hole grey body factors and d-brane spectroscopy}",
    eprint = "hep-th/9609026",
    archivePrefix = "arXiv",
    reportNumber = "RU-96-78",
    doi = "10.1103/PhysRevD.55.861",
    journal = "Phys. Rev. D",
    volume = "55",
    pages = "861--870",
    year = "1997"
}

@article{Aharony:1999ti,
    author = "Aharony, Ofer and Gubser, Steven S. and Maldacena, Juan Martin and Ooguri, Hirosi and Oz, Yaron",
    title = "{Large N field theories, string theory and gravity}",
    eprint = "hep-th/9905111",
    archivePrefix = "arXiv",
    reportNumber = "CERN-TH-99-122, HUTP-99-A027, LBNL-43113, RU-99-18, UCB-PTH-99-16, LBL-43113",
    doi = "10.1016/S0370-1573(99)00083-6",
    journal = "Phys. Rept.",
    volume = "323",
    pages = "183--386",
    year = "2000"
}

@article{Hawking:1975vcx,
    author = "Hawking, S. W.",
    editor = "Gibbons, G. W. and Hawking, S. W.",
    title = "{Particle Creation by Black Holes}",
    doi = "10.1007/BF02345020",
    journal = "Commun. Math. Phys.",
    volume = "43",
    pages = "199--220",
    year = "1975",
    note = "[Erratum: Commun.Math.Phys. 46, 206 (1976)]"
}

@article{Seiberg:1999xz,
    author = "Seiberg, Nathan and Witten, Edward",
    title = "{The D1 / D5 system and singular CFT}",
    eprint = "hep-th/9903224",
    archivePrefix = "arXiv",
    reportNumber = "IASSNS-HEP-99-27",
    doi = "10.1088/1126-6708/1999/04/017",
    journal = "JHEP",
    volume = "04",
    pages = "017",
    year = "1999"
}

@article{Ashok:2021ffx,
    author = "Ashok, Sujay K. and Troost, Jan",
    title = "{Twisted strings in three-dimensional black holes}",
    eprint = "2112.08784",
    archivePrefix = "arXiv",
    primaryClass = "hep-th",
    doi = "10.1140/epjc/s10052-022-10903-y",
    journal = "Eur. Phys. J. C",
    volume = "82",
    number = "10",
    pages = "913",
    year = "2022"
}

@article{Hemming:2001we,
    author = "Hemming, Samuli and Keski-Vakkuri, Esko",
    title = "{The Spectrum of strings on BTZ black holes and spectral flow in the SL(2,R) WZW model}",
    eprint = "hep-th/0110252",
    archivePrefix = "arXiv",
    reportNumber = "HIP-2001-56-TH",
    doi = "10.1016/S0550-3213(02)00021-4",
    journal = "Nucl. Phys. B",
    volume = "626",
    pages = "363--376",
    year = "2002"
}

@article{Massai:2025nci,
    author = "Massai, Stefano and Turetta, Enrico",
    title = "{Three-charge black holes from the worldsheet}",
    eprint = "2511.02499",
    archivePrefix = "arXiv",
    primaryClass = "hep-th",
    month = "11",
    year = "2025"
}

@article{Giveon:2023rsk,
    author = "Giveon, Amit",
    title = "{Negative single-trace $ T\overline{T} $  holography versus de Sitter holography}",
    eprint = "2311.15895",
    archivePrefix = "arXiv",
    primaryClass = "hep-th",
    doi = "10.1007/JHEP05(2024)138",
    journal = "JHEP",
    volume = "05",
    pages = "138",
    year = "2024"
}

@article{Chakraborty:2025rfm,
    author = "Chakraborty, Soumangsu and Mehta, Madhur and Patashuri, Gela",
    title = "{Extremal curves in single-trace $T\overline{T }$-holography}",
    eprint = "2508.09990",
    archivePrefix = "arXiv",
    primaryClass = "hep-th",
    doi = "10.1007/JHEP11(2025)128",
    journal = "JHEP",
    volume = "11",
    pages = "128",
    year = "2025"
}

@article{Chakraborty:2022dgm,
    author = "Chakraborty, Soumangsu and Goykhman, Mikhail",
    title = "{Solvable time-like cosets and holography beyond AdS}",
    eprint = "2204.03024",
    archivePrefix = "arXiv",
    primaryClass = "hep-th",
    doi = "10.1007/JHEP08(2022)244",
    journal = "JHEP",
    volume = "08",
    pages = "244",
    year = "2022"
}

@article{Araujo:2018rho,
    author = "Araujo, Thiago and Colg{\'a}in, E. {\'O} and Sakatani, Yuho and Sheikh-Jabbari, M. M. and Yavartanoo, Hossein",
    title = "{Holographic integration of $T \bar{T}$ {\textbackslash}{\&} $J \bar{T}$ via $O(d,d)$}",
    eprint = "1811.03050",
    archivePrefix = "arXiv",
    primaryClass = "hep-th",
    reportNumber = "IPM/P-2018/076",
    doi = "10.1007/JHEP03(2019)168",
    journal = "JHEP",
    volume = "03",
    pages = "168",
    year = "2019"
}

\end{document}